\documentclass[prd,aps,amsfonts,showpacs,longbibliography,notitlepage,twocolumn,groupedaddress,nofootinbib,superscriptaddress]{revtex4-1}
\usepackage{preamble}

\newcommand{\sout}[1]{ { \color{red} {#1}} }

\begin{document}
\title{
Lifting the mixed to the pure: quantum phases of matter from canonical purification}

\author{Xiaoyang Huang}
\email{xhuang1@perimeterinstitute.ca}
\affiliation{Perimeter Institute for Theoretical Physics, Waterloo, ON N2L 2Y5, Canada}

\author{Shengqi Sang}
\affiliation{Department of Physics, Stanford University, Stanford, California 94305, USA}

\author{Chong Wang}
\affiliation{Perimeter Institute for Theoretical Physics, Waterloo, ON N2L 2Y5, Canada}

\author{Yijian Zou}
\email{yijianzo@amazon.com}
\affiliation{Perimeter Institute for Theoretical Physics, Waterloo, ON N2L 2Y5, Canada}
\affiliation{Amazon Center for Quantum Computing, Pasadena, CA 91125, USA}\thanks{This work was completed prior to joining Amazon.}

\begin{abstract}
  Pure-state quantum phases are defined through finite-depth local unitary connections. This concept of phases has recently been generalized to mixed states in terms of locally reversible channels. In this work, we establish a fundamental relation between mixed-state phases and pure-state phases. We show that a locally reversible channel between mixed states can be lifted to a local isometry between their canonical purifications; thus, if two mixed states are in the same phase, then their canonical purifications are in the same pure-state phase. 
  Furthermore, an explicit construction of the lifting isometry is given by ``half'' of the Petz recovery map, suggesting that the Petz recovery map \emph{without rotation} is enough to connect mixed state phases. We illustrate our results with two examples: toric code under bit flip noise and Chern insulator under fermion number noise.
\end{abstract}

\date{\today}

\maketitle

\emph{Introduction.}--- Classifying phases of matter amounts to identifying many-body states that share the same universal long-distance structure. For pure states, a standard operational formulation uses finite-depth local unitary (FDLU) circuits \cite{HastingsWen2005,ChenGuWen2010}. Extending this formulation to mixed states is subtle: a local quantum channel can destroy information and correlations, and therefore may not preserve the structures that distinguish phases \cite{ZouSangHsieh2023Channeling,FanEtAl2024Diagnostics}. The key to defining mixed-state phases lies in the reversibility of channel actions. This has motivated several notions of mixed-state phase equivalence \cite{CoserPerezGarcia2019,SangZouHsieh2024,MaWang2023AverageSPT,RakovszkyEtAl2024StablePhases,LessaEtAl2025HigherForm}. Meanwhile, substantial progress has been made in understanding intrinsically mixed-state phenomena without direct pure-state counterpart \cite{WangWuWang2025,LessaEtAl2025,MaEtAl2025AverageSymmetries,EllisonCheng2025Classification,LessaChengWang2025Anomaly,HuangEtAl2025Hydrodynamics,Wang2026Review}.

A recent refinement~\cite{SangEtAl2025LocalReversibility,SangHsieh2025,Yi2026} defines phase equivalence using locally reversible channel circuits, so that the locality-preserving character of FDLUs is retained even though the elementary maps are nonunitary \cite{SangEtAl2025LocalReversibility}. Closely related work introduced the Markov length, the scale governing the decay of conditional mutual information across a buffer, and showed that a local noisy evolution can be reversed quasi-locally when the Markov length remains finite along the evolution \cite{SangHsieh2025}. These results connect mixed-state phase stability to local recoverability, but leave open whether the resulting equivalence can be seen in their purifications. Indeed, an arbitrary purification cannot see the mixed-state phase transitions because every local channel admits a Stinespring dilation representation in terms of local unitaries.

In this Letter, we show that mixed-state phase transitions can be probed through their canonical purifications. Defining canonical purification here is distinguished because it is fixed by the mixed-state density matrix up to a basis convention \cite{LuZhuLu2024Bilayer,LuoWangBi2025Holography} and naturally underlies reflected entropy \cite{DuttaFaulkner2021} and fidelity- or R\'enyi-1-based probes of mixed-state order \cite{SalaEtAl2024,LessaEtAl2025,Weinstein2025,ZhangQiBi2026Fidelity,LiuZou2026Fidelity,DiviLessaWang2026LocalSWSSB,LiuYiElse2026,Zhang2026}. Complementary recent work showed that local stability implies clustering of correlations in the canonical purification \cite{LiFirankoHsieh2026}.  The relevant question is therefore more precise: under what dynamical or recoverability assumptions does a local channel acting on a mixed state lift to a local unitary/isometry acting on its canonical purification?

\begin{figure}[t]
  \centering
  \includegraphics[width=0.48\textwidth]{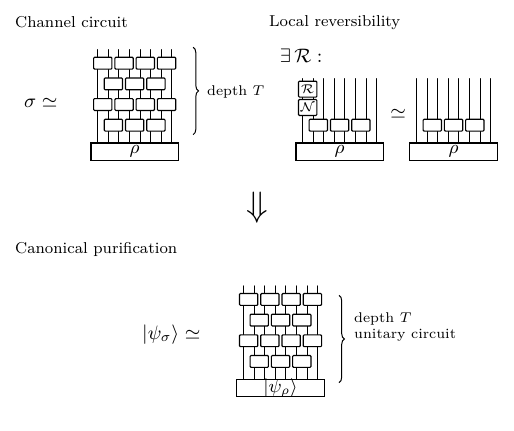}
  \caption{Circuit equivalence. Top left is a depth-$T$ local channel circuit connecting $\rho$ and $\sigma$. Top right is the local reversibility condition. Bottom is a depth-$T$ local unitary circuit connecting their canonical purifications. See the main text for the precise statement. }
  \label{fig:main}
\end{figure}

In this work, we establish such a lifting for locally reversible channels. We start by proving that for an approximate recovery,  a local isometry can approximately connect the canonical purifications with a quatitatively similar error. Such a local isometry is given by ``half'' of the Petz recovery map \cite{hayden_structure_2004,Petz1986Sufficient,Petz1988Sufficiency}. Then, by repeating such construction in a gate-by-gate fashion, we identify a local isometry/unitary circuit that connects two canonical purification states (see Fig.~\ref{fig:main}). We support these results with numerical studies of two representative systems: the toric code under bit flip noise and a Chern insulator subject to fermion number dephasing. Our result partially justifies a body of previous works that used canonical purification~\cite{LiuChenZhangZhouZhang2025,Weinstein2025,LuZhuLu2024Bilayer}, or its less faithful R\'enyi-$2$ proxy (the Choi state)~\cite{LeeYouXu2022,LeeJianXu2023}, to characterize and classify mixed-state phases from pure-state physics~\cite{Ma2024SPTdoubled,EllisonCheng2025Classification,SohalPrem2025,Zhang2025,SymTFT1,SymTFT2,LuoWangBi2025Holography,ManjunathEtAl2025Detection}.

\emph{Recovery for canonical purifications}---
Given a density matrix $\rho$ on a finite-dimensional Hilbert space $\mathcal{H}$, the canonical purification is defined by vectorizing the operator $\sqrt{\rho}$:
\begin{align}
    |\psi_\rho\rangle : = \sqrt{\rho} \otimes I_{\bar{\mathcal{H}}}|\Omega\rangle,\quad |\Omega\rangle := \sum_s |s\rangle \otimes |s\rangle,
\end{align}
where $|\Omega\rangle$ is the unnormalized maximally entangled state between $\mathcal{H}$ and its conjugate Hilbert space $\bar{\mathcal{H}}$. 
% Let $\rho_{AB}$ be a bipartite density matrix, and let $\mathcal{N}_{B\to C}$ be a quantum channel such that $\mathcal{N}_{B\to C}(\rho_{AB}) = \sigma_{AC}$. 
Consider two density matrices $\rho_{AB}$ and $\sigma_{AC} := \mathcal{N}_{B\to C}(\rho_{AB})$, where $\mathcal{N}_{B\to C}$ is a quantum channel acting on subsystem $B$ \footnote{For simplicity, we assume that all density matrices are full rank throughout the proof. This is not an essential restriction in finite dimensions, as the singular case follows by restricting to the relevant supports and taking a full-rank approximation, with the result recovered by continuity. }
Denote by $|\psi_\rho\rangle_{AA'BB'}$ and $|\psi_\sigma\rangle_{AA'CC'}$ the corresponding canonical purifications, where primed labels indicate the conjugate copies of the associated regions. We are interested in the local reversibility (LR) of the channel, i.e., whether there exists a channel $\mathcal{R}_{C\rightarrow B}$ such that $\mathcal{R}(\sigma_{AC}) = \rho_{AB}$. It is well known that LR is equivalent to the non-decreasing of mutual information: $I(A:B)_{\rho} = I(A:C)_{\sigma}$, where $I(A:B)_{\rho}:= S(\rho_{A}) + S(\rho_B) - S(\rho_{AB})$ \cite{hayden_structure_2004,Petz1986Sufficient,Petz1988Sufficiency}. In a similar manner, we define LR for the canonical purification as there existing an isometry $W_{CC'\to BB'}$ (assuming $d_B>d_C$) such that, 
\begin{equation}
    |\psi_\rho\rangle_{AA'BB'} =W_{CC'\to BB'}  |\psi_\sigma\rangle_{AA'CC'}.
\end{equation}

The key step towards our main results is given by the following lemma, which is based on the results in \cite{GaoWilde2021Recoverability}. For a tripartite state $\rho = \rho_{ABC}$, set $\eta:= I(A:C|B)_\rho = I(A:BC)_\rho - I(A:C)_\rho$ which is the conditional mutual information.
\begin{lem}\label{lem:main}
    For every tripartite state with  $0<\eta\le1$,
\begin{equation}
\norm{\sqrt{\rho} - \rho_{AB}^{1/2} \rho_B^{-1/2} \rho_{BC}^{1/2} }_2\le\frac{\sqrt\eta}{\pi}
\left(2+\sqrt{2\log\frac{16d_*}{\eta}}\right),
\label{eq:main}
\end{equation}
where $\norm{X}_2:=\sqrt{\tr(X^\dagger X)}$, and $d_\star:= \min\{d_A,d_C\}$ is the minimum Hilbert-space dimension of regions $A,C$.
\end{lem}
\noindent The proof of Lemma \ref{lem:main} is given in Appendix.\ref{app:ordered-petz}. Observe that the approximation to $\sqrt{\rho}$ is given by ``half'' of the Petz map associated with channel $\mathcal{N}=\tr_C$:
\begin{align}
    \mathcal{R}_{\rho_{BC},\mathcal{N}} (\cdot):= \rho_{BC}^{1/2}\rho_B^{-1/2}(\cdot) \rho_B^{-1/2} \rho_{BC}^{1/2}.
\end{align}
Indeed, using  $\norm{XY}_1\leq \norm{X}_2\norm{Y}_2$ and $\norm{\sqrt{\rho}}_2 =\norm{ \rho_{AB}^{1/2} \rho_B^{-1/2} \rho_{BC}^{1/2}}_2=1$, we have
\begin{align}\label{eq:main approximate recovery Petz}
    \norm{\rho - \mathcal{R}_{\rho_{BC},\mathcal{N}} (\rho_{AB}) }_1\le\frac{\sqrt\eta}{\pi}
\left(2+\sqrt{2\log\frac{16d_*}{\eta}}\right).
\end{align}
Remarkably, this bound on the Petz map has not appeared in the literature to our knowledge. In particular, previous bounds on the Petz map recovery all involve the operator norm of the inverse of the total density matrix \cite{wilde_recoverability_2018,CarlenVershynina2018QuasiEntropies,CarlenVershynina2020Recovery}, which is generally exponentially large in many-body systems. By contrast, \eqref{eq:main approximate recovery Petz} gives a tighter bound that involves the dimension of subregions inside a logarithm, which is mostly polynomial in system size. 

The lemma can be reinterpreted as a relation between LR of pure and mixed states (for the special case of $\mathcal{N}=\tr_C$), which is one of our main results:
\begin{thm}\label{thm:main petz}
   Given a tripartite state $\rho_{ABC}$ satisfying $1>\eta\geq0$, define the isometry:
   \begin{align}
       W_{B\rightarrow BCC'} := \rho^{1/2}_{BC}\left( \rho^{-1/2}_{B}\otimes I_{CC'}\right)(I_B \otimes \ket{\Omega_{CC'}}),
   \end{align}
   with $W^\dagger W = I_B$.
   % where $\ket{\Omega_{CC'}}=\sum_i \ket{i}_C\ket{i}_{C'}$ is the unormalized maximally entangled state between $C$ and $C'$, 
   Then it is satisfied that:
   \begin{align}\label{eq:bound petz}
       \abs{\langle\psi_\rho|W_{B\to BCC'} |\psi_{\rho_{AB}}\rangle}\geq 1- \delta(\eta),
   \end{align}
   with
   \begin{align}
       \delta(\eta):= \frac{\eta}{2\pi^2} \left(2+\sqrt{2\log\frac{16d_*}{\eta}}\right)^2.
   \end{align}
   If $\eta = 0$, then $\delta(0)=0$. 
\end{thm}

For the case of exact LR, $\eta=0$, we have an exact isometry that connects the canonical purifications. This can also be seen using the structure theorem for exact Markov chains~\cite{hayden_structure_2004}. Specifically, a tripartite state has a decomposition $\rho_{ABC} = \sum_{i} p_i \rho_{AB^{i}_L} \otimes \rho_{B^i_R C}$. By orthogonality of the subspaces, we have $\sqrt{\rho_{ABC}} =  \rho_{AB}^{1/2} \rho_B^{-1/2} \rho_{BC}^{1/2}$, which gives the exact isometry; see Appendix.~\ref{app: exact} for a full proof. Furthermore, the exact isometry conversely implies the existence of an exact LR channel, and we will comment on this result in the discussion. 

For applications to many-body physics, the exact LR is often weakened to the approximate LR. 
For the class of commuting mixed states, including classical distributions and stabilizer states under Pauli noise, being an approximate quantum Markov chain is close
to being an exact quantum Markov chain. Therefore, the equivalence of the approximate LR in these cases becomes a corollary of the exact LR results; see Appendix~\ref{app:approx classical}. 
For generic quantum states, however, approximate Markovianity does not imply closeness to an exact quantum Markov chain --- the antisymmetric state is a standard example~\cite{christandl_entanglement_2012}. As a sanity check, we show in Appendix~\ref{app:antisymmetric state} that the approximate LR still holds for the canonical purifications of the antisymmetric states.

Theorem \ref{thm:main petz} can be generalized to an arbitrary quantum channel using the Stinespring dilation theorem. Consider an isometry $V_{B\to CE}$ of the channel $\mathcal{N}_{B\to C}$ such that $\sigma_{AC} = \tr_E(\gamma_{ACE})$ with $\gamma_{ACE}:= V_{B\to CE} \rho_{AB} V^\dagger_{B\to CE}$. Let
\begin{align}
    \epsilon:= I(A:B)_\rho - I(A:C)_\sigma = I(A:E|C)_\gamma.
\end{align}
Then Theorem \ref{thm:main petz} implies $|\psi_\rho\rangle$ and $|\psi_\sigma\rangle$ are also related by an isometry $\tilde W_{CC'\to BB'}$ with error up to $\delta(\epsilon)$.

Moreover, we can improve the bound to be dimension-independent. The following theorem is based on the $p$-norm generalization of the universal recovery theorem with the ``rotated'' Petz map~\cite{junge_multivariate_2022}.
\begin{theoremprime}\label{thm:main rotated petz}
    Given a state $\rho_{AB}$ and a channel $\mathcal{N}_{B\to C}$. Let $\sigma_{AC} := \mathcal{N}_{B\to C}(\rho_{AB})$. Then, there exists an isometry $W_{CC'\to BB'}$ such that
   \begin{align}\label{eq:bound rotated petz}
       \abs{\langle\psi_\rho|W_{CC'\to BB'} |\psi_\sigma\rangle}\geq e^{-\epsilon/2}.
   \end{align}
\end{theoremprime}
\noindent The proof is given in Appendix~\ref{app:approx quantum}; see v2 of \cite{LiFirankoHsieh2026} for a similar result.

As soon as we require $\epsilon$ to decay exponentially in linear size of $B$ and $C$, the influence of the dimension factor in \eqref{eq:bound petz} will be negligible. Consequently, the bounds in theorem \ref{thm:main petz} and \ref{thm:main rotated petz} are equally well suited for applications in this regime. Such a regime is relevant to the following discussions about many-body phases.

\textit{Mixed-state phases. --} We now apply Theorem \ref{thm:main petz}/\ref{thm:main rotated petz} to mixed state phases. We first give an informal definition of mixed state equivalence:

\begin{defn}[modified Def.3 in Ref.\cite{MaKhemaniSang2026FiniteTemperature}]\label{def:main}
Let $\mathcal{C}$ denote a local channel circuit given by
\begin{align}
    \mathcal{C}=\mathcal{C}_T \circ \cdots \circ \mathcal{C}_2 \circ \mathcal{C}_1 \qquad  \mathcal{C}_t=\prod_x \mathcal{E}_{t, x}
\end{align}
where different gates $\mathcal{E}_{t, x}$ in a
layer $t$ have finite non-overlapping spatial support.
$\mathcal{C}$ is $\varepsilon$-locally reversible with respect to an input state $\rho$, if, for each gate $\mathcal{E}_{t,x}$, there exists another channel $\tilde{\mathcal{E}}_{t,x}$ with the same spatial support, such that
\begin{align}\label{eq:def LR}
    \norm{\tilde{\mathcal{E}}_{t,x} \mathcal{E}_{t,x} \mathcal{C}'[\rho] - \mathcal{C}'[\rho] }_1\leq \varepsilon,
\end{align}
where $\mathcal{C}':= \mathcal{C}_{t-1}\cdots \mathcal{C}_1$. 

If the error $\varepsilon$ decays faster than any polynomial function of system size $L$, $\varepsilon \cdot \mathrm{Poly}(L)\ll 1$, we say $\mathcal{C}$ is a LR channel circuit.

If two mixed states $\rho$ and $\sigma$ are connected by a finite-depth/finite-layer LR channel circuit, we say that they are in the same mixed-state phase.
\end{defn}
\noindent Intuitively, Definition~\ref{def:main} guarantees that after applying a partial channel circuit, every intermediate step is also LR. Therefore, a channel circuit $\tilde{\mathcal{C}}$ made of $\tilde{\mathcal{E}}_{t,x}$ recovers the state with
\begin{align}
    \norm{\tilde{\mathcal{C}}\mathcal{C}[\rho] - \rho }_1 &\leq \sum_{t=1}^T \sum_x \norm{\tilde{\mathcal{E}}_{t,x} \mathcal{E}_{t,x} \mathcal{C}'[\rho] - \mathcal{C}'[\rho] }_1 \nonumber\\
    &\leq  \varepsilon  T\cdot \mathrm{Poly}(L) \to 0
\end{align}
A direct consequence of Theorem~\ref{thm:main petz}/\ref{thm:main rotated petz} is that LR channel circuits lift to local unitary circuits on canonical purifications: 

\begin{cor}
\label{main:cor}
    If two density matrices $\rho$ and $\sigma$ are in the same mixed-state phase, then their canonical purifications are in the same pure-state phase, i.e., they are approximately connected by an FDLU.
\end{cor}
\begin{proof}
    For each state $\rho'=\mathcal{C}'[\rho]$ and gate $\mathcal{E}_{t,x}$ with input support $B$ and output support $C$ satisfying \eqref{eq:def LR} with error $\varepsilon$, the standard entropy analysis gives \cite{FawziRenner2015ApproximateMarkov}
    \begin{align}
        \epsilon \lesssim \varepsilon \log\varepsilon^{-1} \log d_B.
    \end{align}
     According to Theorem~\ref{thm:main petz}/\ref{thm:main rotated petz}, there exists an isometry $W_{t,x}$ from $CC'\to BB'$ such that $\abs{\langle\psi_{\rho'}|W_{t,x} |\psi_{\sigma'}\rangle}\geq 1- \delta(\epsilon)$ where $\sigma'= \mathcal{E}_{t,x} (\rho')$. Extending the isometry to a unitary $U_{t,x}$ will not change the overlap. Therefore, by defining $\mathcal{U}= \prod_{t,x}U_{t,x}$, we have
     \begin{align}
         \abs{\langle \psi_\rho|\mathcal{U}|\psi_\sigma\rangle} \gtrsim 1- \varepsilon T  \cdot \mathrm{Polylog}(\varepsilon^{-1})   \mathrm{Poly}(L) \to 1,
     \end{align}
     where we used $\log d_B,\log d_C\lesssim \mathrm{Poly}(L)$. By construction, $\mathcal{U}$ is a FDLU.
\end{proof}
\noindent 
An illustrating figure of Corollary \ref{main:cor} is shown in Fig.\ref{fig:main}.

The results above suggest that canonical purification can be used to detect mixed-state phase transitions driven by quantum channels. A sufficient condition for the LR channel circuit is that every intermediate state has finite Markov length \cite{SangHsieh2025,SangEtAl2025LocalReversibility}.
Below, we demonstrate how finite Markov length implies the same phase of canonical purification with two examples. 

% \yijian{Two examples should be written as two different sections...} \yijian{A good idea is to draw phase diagrams of the two models, both the mixed-state phases and the canonical purification phases.}\XYH{sure. maybe we could put it in the beginning?}

 % The logarithmic dimension factor $\log d_B$ comes from continuity bounds for the entropy and conditional entropy (see Appendix~\ref{app:proof}). It is harmless for the circuit-based phase equivalence because it will be absorbed into a factor of $\mathrm{Poly}(L)$ with $L$ the linear system size; as we will see, quasi-locality requires $\epsilon \cdot \mathrm{Poly}(L)\ll 1$.

\emph{Toric code under bit flip noise.}---The first example is the toric-code ground state subjected to the dephasing channel $\mathcal{E}_p(\rho):= (1-p)\rho + p X \rho X, p\in[0,1/2]$, where $X$ is the Pauli-$X$ operator. The ground state $|\psi_{TC}\rangle$ is a stabilizer state with Pauli-$Z$ plaquette and Pauli-$X$ vertex stabilizers on a $L\times L$ square lattice. After the channel is applied, the mixed state $\rho_p$ is a mixture of different $m$-anyon excitations. It is well established that $\rho_p$ undergoes a decoding transition at $p_c\approx 0.109$~\cite{noauthor_topological_nodate}, and later Ref.~\cite{SangHsieh2025} proposed that the same $p_c$ also marks the mixed-state phase transition: the CMI remains short-ranged for $p<p_c$ and $p>p_c$, but becomes long-ranged at $p=p_c$. Thus, mixed states on the same side of $p_c$ can be connected by LR channels, whereas states on opposite sides cannot \cite{SangHsieh2025,SangEtAl2025LocalReversibility}. Based on Theorem~\ref{thm:main petz}/\ref{thm:main rotated petz}, we conclude the following results. For $p<p_c$, the canonical purification $|\psi_{\rho_p}\rangle$ is connected by FDLU to the $p=0$ limit $|\psi_{TC}\rangle \otimes |\psi_{TC}\rangle$, which consists of two copies of toric code.  For $p>p_c$, the canonical purification is connected to the $p=1/2$ limit, which is given by $\prod_i  \mathrm{CNOT}_i \left(|\psi_{TC}\rangle \otimes |+\rangle^{\otimes 2L^2} \right)$ where $\mathrm{CNOT}_i$ controls on qubit $i$ of the second copy and acts on qubit $i$ of the first copy; hence, it has the same topological order as the single copy of toric code.

\begin{figure}[tp]
  \centering
  \includegraphics[width=0.48\textwidth]{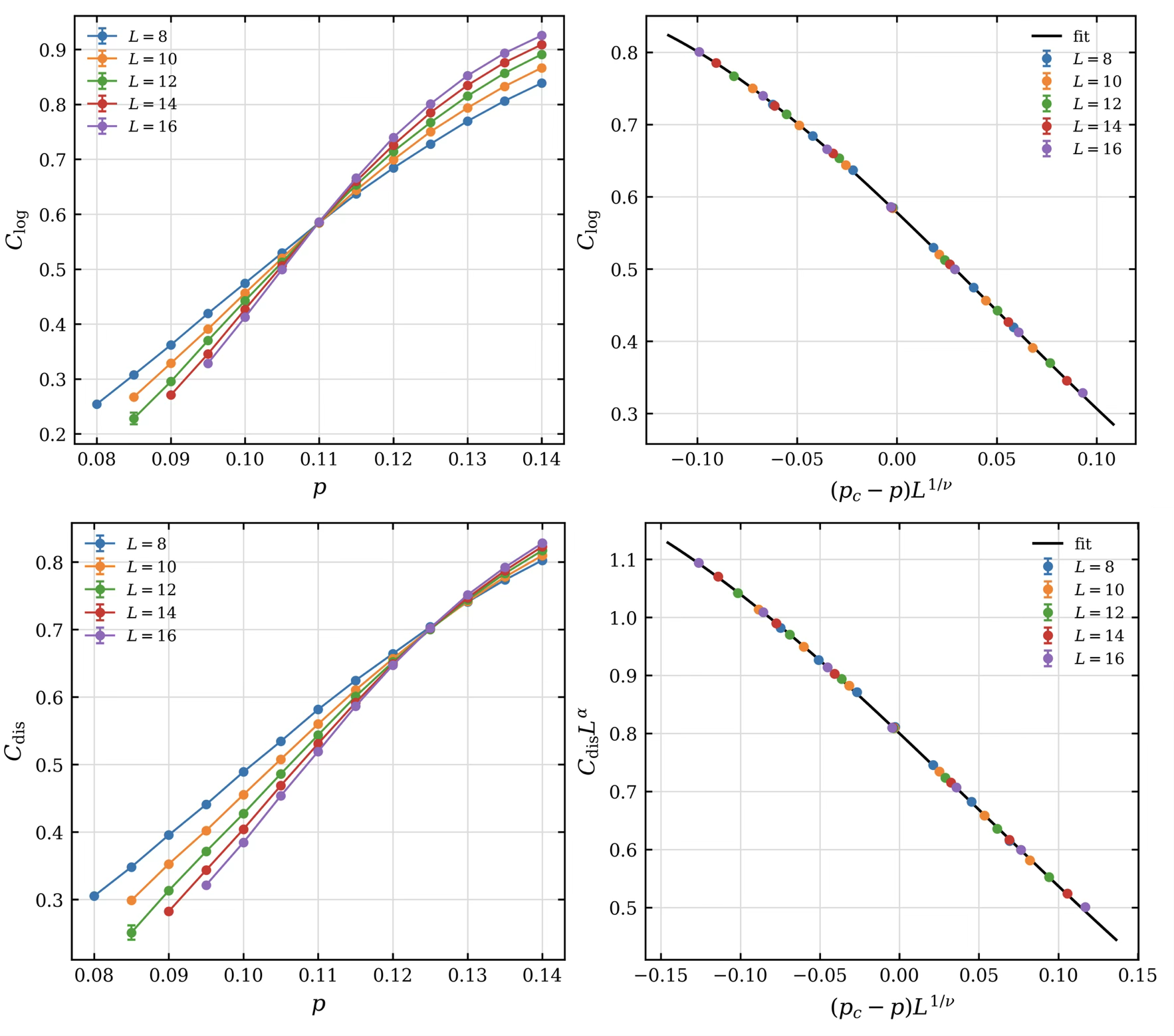}
  \caption{Expectation values of string operators in the canonical purification of the dephased toric-code state. Top: the logical operator jumps as $p$ increases. A data collapse over system size gives $p_c\approx 0.109$ and $\nu_{\mathrm{log}}\approx 1.493$. Bottom: the disorder operator jumps as $p$ increases. After including the endpoint scaling dimension $\alpha \approx 1.6$, the curves collapse with $p_c\approx 0.109$ and $\nu_{\mathrm{dis}}\approx 1.325$. The data are averaged over $10^6$ samples, and most error bars are smaller than the markers. The discrepancy in $\nu_{\mathrm{dis}}$ is due to finite-size effects.}
  \label{fig:string}
\end{figure}

To confirm that the same $p_c$ also characterizes the phase transition of the canonical purifications $|\psi_{\rho_p}\rangle$, we study correlation functions in the doubled Hilbert space. Specifically, we use an open operator, consisting of an open string with two endpoints, and a logical operator, consisting of a noncontractible loop:
\begin{align}
    C_{\mathrm{dis}}:=& \langle\psi_{\rho_p}| \prod_{i\in \gamma_{\mathrm{open}}}(X_i \otimes X_i) |\psi_{\rho_p} \rangle \\
    C_{\mathrm{logical}}:=& \langle\psi_{\rho_p}| \prod_{i\in \gamma_{\text{logical}}}(X_i \otimes X_i) |\psi_{\rho_p} \rangle, 
\end{align}
where $\gamma$ is a path in the dual lattice and $i\in \gamma$ denotes the edge $e_i$ traversed by $\gamma$. We take $|\gamma_{\mathrm{open}}| = L/2$. The numerics is performed up to $L=16$ based on mapping to the two-dimensional random-bond Ising model (RBIM)~\cite{noauthor_topological_nodate}, which acquires a free-fermion representation; see Appendix~\ref{app:toric code}.

As shown in Fig.~\ref{fig:string}, $C_{\mathrm{logical}}$ decays to zero for $p<p_c$ while remains finite for $p>p_c$. This can be easily verified at the extreme points: at $p=0$, $\langle\psi_{TC}|X_i|\psi_{TC}\rangle = 0$; at $p=1/2$, $\mathrm{CNOT}_i(X_i\otimes X_i) \mathrm{CNOT}_i = \mathbf{1}\otimes X_i$ and $\langle+|X_i|+\rangle = 1$. Thus, the numerical results are consistent with our expectations that $\lim_{L\to\infty}C_{\mathrm{logical}}$ should not change under FDLU; in fact, $C_{\mathrm{logical}}$ is the Holevo fidelity that encodes the decoding success (see \cite{JouzdaniNovaisMucciolo2013SurfaceCodeBosonicBath,LessaEtAl2025} for related measure).
Unlike the logical operator, the disorder operator carries $m$-anyons at its two endpoints, which have a nontrivial scaling dimension. We therefore characterize the transition using the scaling ansatz
\begin{align}
    C_{\mathrm{dis}} &\propto L^{-\alpha}f_1((p_c - p) L^{1/\nu_{\mathrm{dis}}}) \\
    C_{\mathrm{logical}} &\propto f_2((p_c - p) L^{1/\nu_{\mathrm{log}}}),
\end{align}
where $f_{1,2}$ are bounded smooth functions. Fixing $\alpha \approx 1.6$ as reported in \cite{merz_negative_2002}, we find that both data sets collapse according to this ansatz with the same $p_c\approx 0.109$ and with exponent close to the reported value in \cite{merz_two-dimensional_2002}.

We also compute the two-point correlation function
\begin{align}
    C_{XX}(r):=&\langle\psi_{\rho_p}| (X_i \otimes X_i)(X_j \otimes X_j) |\psi_{\rho_p} \rangle \nonumber \\
    &- \langle\psi_{\rho_p}| X_i \otimes X_i |\psi_{\rho_p} \rangle \langle\psi_{\rho_p}| X_j \otimes X_j |\psi_{\rho_p} \rangle,
\end{align}
where $r = |i-j|$. Since $C_{XX} = 0$ in both the clean limit and the maximally dephased limit, by Lieb-Robinson bound, we expect $C_{XX}(r)$ to decay exponentially away from the critical point. Near the critical point, we find a power-law decay of $C_{XX}(r)$ as shown in Appendix~\ref{app:toric code}.

\emph{Chern insulator under flagged dephasing.}---
Our second example is a Chern insulator subjected to flagged fermion-number dephasing. On each site, we consider the dephasing channel
    $\mathcal{N}^{[i]}_{p}(\rho)
    =
    \left(1-\frac{p}{2}\right)\rho
    +\frac{p}{2}(1-2n_i)\rho(1-2n_i)$,
where \(n_i=c_i^\dagger c_i\). We use its flagged realization
\begin{align}
    \mathcal{N}^{[i]}_{p,F}(\rho)
    ={}&(1-p)\rho\otimes\ket{0}\!\bra{0}_{F}
    +p\,\mathcal{N}^{[i]}_{1}(\rho)\otimes\ket{1}\!\bra{1}_{F},
    \label{eq:flagged-dephasing}
\end{align}
whose flag trace reproduces \(\mathcal{N}^{[i]}_{p}\). As opposed to the decohered toric code, the state is not stabilizer state and the marginal density matrices do not commute. 

The state $\mathcal{N}^{[i]}_{p,F}(\rho)$ is locally reversibly connected to the measurement ensemble\footnote{Since the complete dephasing is equivalent to the projective measurement of fermion occupation number, we can recover the measurement outcome by measuring $n_i$ conditioned on the dephasing flag being one; this operation does not collapse the state as the state is already diagonal in occupation basis. Conversely, the dephasing flag is recovered by retaining whether a measurement is performed.}. Therefore we can expand the decohered states into measurement trajectories.
\begin{align}
    \rho_{p,F}
    =\sum_{\mathbf m}p_{\mathbf m}\rho_{\mathbf m}
      \otimes\ket{\mathbf m}\!\bra{\mathbf m}_{F},
\end{align}
where we apply independent $n_i$ measurement on each site $i$ with probability $p$, with the flag $|\mathbf{m}\rangle$ recording the measurement outcomes and $\rho_{\mathbf{m}}$ the post-measurement state.
We will consider the mixed-state phase of $\rho_{p,F}$ and the pure-state phase of its canonical purification. Each state $\rho_{\mathbf{m}}$ is a Gaussian state, which enables us to evaluate physical quantities efficiently, see appendix for details.

\begin{figure}[t]
    \centering
    \begin{subfigure}[t]{0.48\columnwidth}
        \centering
        \includegraphics[width=\linewidth]{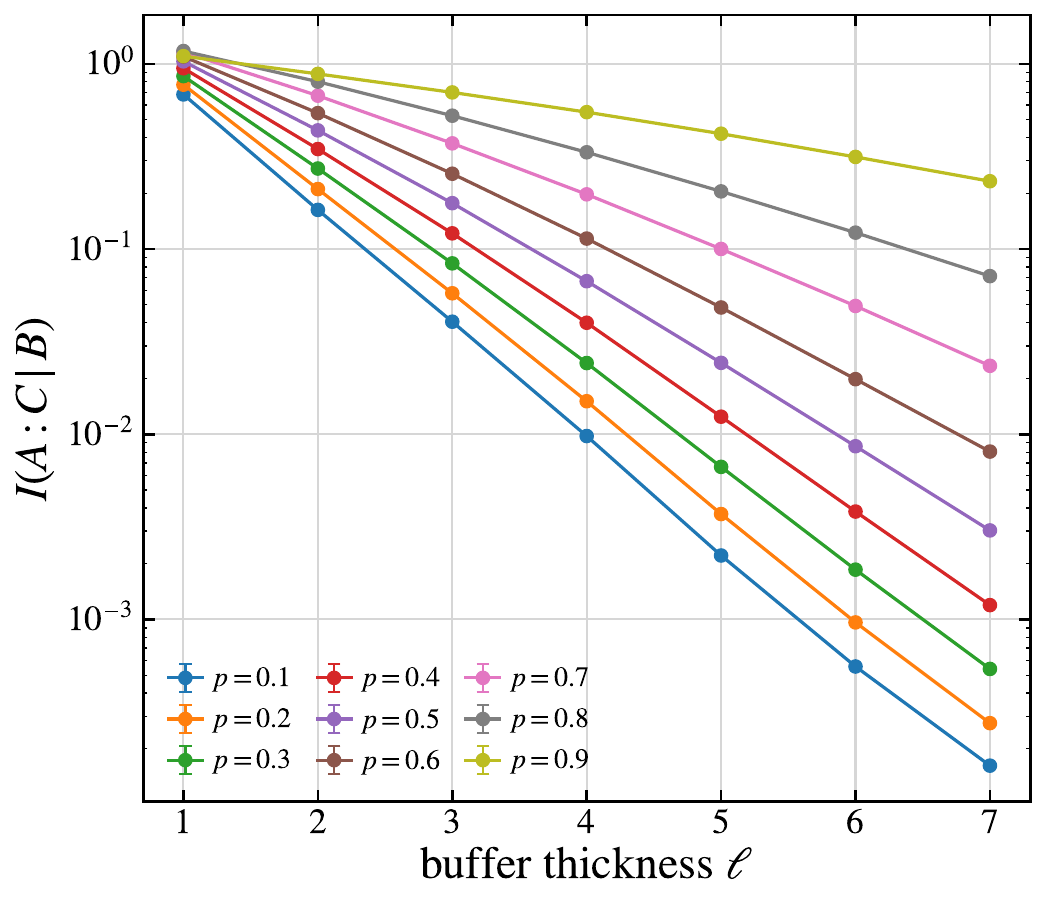}
        \caption{}
    \end{subfigure}
    \hfill
    \begin{subfigure}[t]{0.48\columnwidth}
        \centering
        \includegraphics[width=\linewidth]{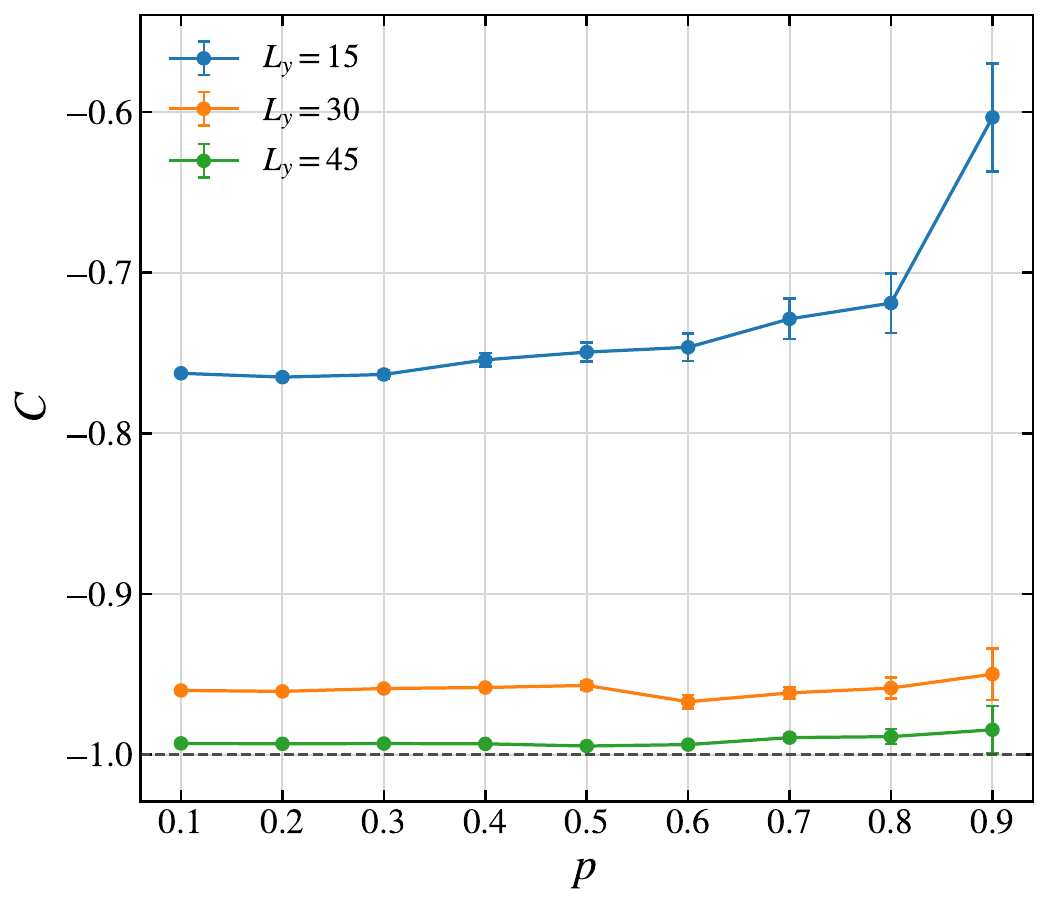}
        \caption{}
    \end{subfigure}
    \caption{
    Chern insulator under flagged fermion-number dephasing.
    (a) Conditional mutual information as a function of buffer thickness
    \(\ell\), shown on a linear--logarithmic scale. The calculation uses a
    \(24\times24\) Hofstadter model with flux \(2\pi/4\), and each point is
    averaged over \(10^4\) trajectories. 
    (b) The spin-Hall Chern number of the canonical purification extracted from
    Eq.~\eqref{eq:chern-from-partial-translation} on \(20\times L_y\)
    cylinders. 
    }
    \label{fig:chern-dephasing}
\end{figure}

We first show that the Markov length remains finite for $p<1$ by numerically computing the CMI \(I(A:C|B)\). We take \(A\) to be a centered \(2\times2\) square, \(B\) to be an annular buffer of thickness \(\ell\), and \(C\) to be the complement of \(AB\). As shown in Fig.~\ref{fig:chern-dephasing}(a), the CMI decays exponentially with \(\ell\) for all \(p=0.1,\ldots,0.9\) and becomes slower as \(p\) approaches 1, indicating no mixed-state phase transitions for the whole $p<1$ region. 

We next examine the canonical purification
\(\ket{\psi_p}=\ket{\sqrt{\rho_{p,F}}}\). The state has $U(1)\times U(1)$ symmetry on the two copies and is time-reversal symmetric. Therefore, the total Chern number vanishes and the nontrivial phase is characterized by nonzero spin-Hall conductance. We diagnose the spin-Hall response using the partial-translation construction \cite{Kobayashi_2025}.
We place the Hofstadter model, with flux $2\pi/5$ and filled lowest band on an \(L_x\times L_y\) open cylinder with antiperiodic boundary conditions around \(y\), and let \(A\) denote its left half. The spin-Hall response is extracted through the expectation value 
\begin{align}
    \mathcal{T}_{s}(p)
    =
    \bra{\psi_p}
    U_{A,\pi/3}U_{A',\pi}
    T_{AA',2\pi/3}
    \ket{\psi_p},
    \label{eq:spin-partial-translation}
\end{align}
where \(U_{A,\phi} = \exp(i \phi\sum_{(x,y)\in A } n_{x,y})\) is a charge rotation, and \(T_{AA',2\pi/3}\) translates both the fermions and the flags in \(A\) and \(A'\) by \(L_y/3\). We generalize the result of Ref.~\cite{Kobayashi_2025} and compute the spin-Hall conductance by
\begin{align}
    C
    =-\frac{3}{2\pi}\operatorname{arg}\mathcal{T}_{s}(p).
    \label{eq:chern-from-partial-translation}
\end{align}

Figure~\ref{fig:chern-dephasing}(b) shows the computed value of $C$ for
\(L_x=20\) and \(L_y=15,30,45\). The finite-circumference corrections decrease systematically with increasing \(L_y\), and the result approaches the quantized value \(C=-1\) throughout the entire sampled interval \(0.1\leq p\leq0.9\). Thus, the canonical purification also remains as a spin-Hall insulator throughout the $p<1$ region. For $p=1$, we further verify that the CMI becomes power-law decay and the canonical purification has power-law decaying two-point correlation functions, indicating criticality for both states. The example thus demonstrates Corollary \ref{main:cor} beyond the commuting case.

\emph{Discussion.}---
\emph{i).} A recent work \cite{LiFirankoHsieh2026} also studied canonical purifications of mixed states that have a small CMI. They focused on the clustering property of the canonical purifications. In particular, they showed that when the mixed state has both decay of CMI and MI, the canonical purification has clustering of local operators. In comparison, our results implies that if a mixed state is in the same mixed-state phase as a trivial product state, then its canonical purification is connected by FDLU to a trivial product state. It would be interesting to investigate more on the relationship with \cite{LiFirankoHsieh2026}.

%\emph{ii).} A well-studied class of mixed-state phase transitions is between classical distributions \cite{ChenGrover2026RGFlows,ChenGrover2026DirectedPercolation, LloydAbaninGopalakrishnan2025ConditionalCorrelation,ZhangGopalakrishnan2025DecoheredGibbs}. %Specifically, consider the %partial-trace channel $\mathcal{N} = \tr_C$, so that %$\mathcal{N}(p_{ABC}) = p_{AB}$. For each configuration $b$ in r%egion $B$ with $p_B>0$, define
%begin{align}
%    \ket{w_b}_{CC'}
%  :=
%  \sum_c \sqrt{p_{C|B}}\,\ket{c}_C\ket{c}_{C'}.
%\end{align}
%There always exists a unitary $U_b$ on $CC'$ such that %$U_b\ket{0}_{CC'}=\ket{w_b}_{CC'}$, since both states are %normalized. Now set
%\begin{align}\label{eq:B-controlled unitary}
%    U_{BB'CC'}
%  =
%  \sum_b \ket{b}\!\bra{b}_B\otimes\ket{b}\!\bra{b}_{B'}\otimes U_b.
%\end{align}
%Then Lemma \ref{lem:main} implies $U_{BB'CC'}\bigl(\ket{\psi_{p_{AB}}}\otimes \ket{0}_{CC'}\bigr) \approx \ket{\psi_{p_{ABC}}}$. %Note that keeping the double copies is not necessary for practical purposes. A detailed analysis of the tensor-network update and truncation will be in an upcoming work 
%\XYH{Yijian: Do you want me to write in this way?}
%\yijian{Seems that we don't have enough space to discuss the simulation aspects. Furthermore, since we now have Lemma 1, the isometry becomes explicit for both classical and quantum cases, so there is no need mentioning classical cases anymore.}

\emph{ii).} Our main result is the phase equivalence between canonical purification for the same mixed-state phase. One natural question is the reverse: given canonical purifications $|\psi_1\rangle$ of $\rho_1$ and $|\psi_2\rangle$ of $\rho_2$, if $|\psi_2\rangle = \mathcal{U}|\psi_1\rangle$ for an FDLU circuit $\mathcal{U}$, does $\rho_1$ and $\rho_2$ belong to the same phase? The answer is negative without further constraints. For a counterexample, consider an $N$-qubit state $\rho_1 = \frac{1}{2}(|0\rangle\langle0|^{\otimes N}+|1\rangle\langle1|^{\otimes N})$ whose canonical purification is a GHZ state: $|\psi_1\rangle = \frac{1}{\sqrt{2}} (|\bar 0\rangle^{\otimes N}+ |\bar 1\rangle^{\otimes N})$ with $|\bar 0\rangle:= | 0\rangle| 0\rangle, |\bar 1\rangle:= | 1\rangle| 1\rangle$. Applying the Hadamard gate to each qubit, we have $|\psi_2\rangle = \prod_i \bar H_i |\psi_1\rangle = 2^{-(N-1)/2} \sum_{\abs{x}\in \mathrm{even}}|x\rangle|x\rangle$ with $x$ a size-$N$ bitstring, and thus $\rho_2 = 2^{-(N-1)} \sum_{\abs{x}\in \mathrm{even}}|x\rangle\langle x| \propto \mathbf{1}+\prod_i Z_i$. However, $\rho_2$ has infinite Markov length \footnote{ It is the parity state in Ref.\cite{LessaEtAl2025} in a different basis.} while $\rho_1$ has zero CMI, so they are not in the same mixed-state phase.
Another counterexample has been given in Ref.~\cite{EllisonEtAl2026}.
It would be interesting to study under which circumstances FDLU connection between canonical purifications imply equivalence of mixed-state phases, perhaps following a reverse theorem of Theorem 2 \cite{LiFirankoHsieh2026}, which we prove and explore independently in the appendix.~\ref{app: exact} 

%\yijian{However, if we further assume that the pure state remains a canonical purification after acting with \textit{any subset of gates} in $\mathcal{U}$ on $|\psi_1\rangle$, then we can prove that the mixed states are in the same phase. This follows from a reverse theorem of Theorem 2 in the exact case \cite{LiFirankoHsieh2026}, which we independently prove in the appendix. }

%For approximate LR, we find that
%two mixed states that are in different mixed-state phases can have their canonical purifications be connected by FDLU.  %Therefore, these examples disprove the reversed direction of Theorem.\ref{thm:main petz}.

\emph{Acknowledgement.}--- We acknowledge useful discussions with Dominic Else. We thank Zhi Li, Raz Firanko, and Timothy H. Hsieh for helpful discussions and for coordinating their related work with ours. S.S. would like to especially thank Chao Yin for discussions, which inspired Lemma 1. OpenAI GPT 5.6 and 6 were used to code and prototype the results, and all the outputs were verified by the authors. S.S. was supported by the SITP postdoctoral fellowship at Stanford University. Research at Perimeter Institute is supported in part by the Government of Canada through the Department of Innovation, Science and Economic Development and by the Province of Ontario through the Ministry of Colleges and Universities. 

\bibliography{canonical}

\newpage

\onecolumngrid
\appendix
\renewcommand{\thesubsection}{\Alph{section}.\arabic{subsection}}

\section{Proofs of the main results}\label{app:proof}

\subsection{Ordered Petz amplitude bound}\label{app:ordered-petz}

We prove Lemma~\ref{lem:main} using Gao--Wilde's \cite{GaoWilde2021Recoverability} Lemmas~4.10 and~4.11.
We first recall the notation needed to state their results, restricted to the zero-rotation case $t=0$ used below.

\paragraph{Notation.}
% We write $S(\omega)=-\operatorname{Tr}(\omega\log\omega)$,
% $D(\omega\Vert\sigma)=\operatorname{Tr}[\omega(\log\omega-\log\sigma)]$.
% Operators on subsystems are extended by identities on the missing factors.
% For singular states, inverse powers mean inverse powers on the support, extended by zero on the kernel.

For the two lemmas stated below, let $\omega,\sigma$ be density operators on $R E$, and set $\omega_R=\operatorname{Tr}_E\omega$ and $\sigma_R=\operatorname{Tr}_E\sigma$. Introduce the matrix algebras $\mathcal M=\mathcal B(\mathcal H_R\otimes\mathcal H_E)$ and $\mathcal N=\mathcal B(\mathcal H_R)\otimes I_E$, with conditional expectation defined as
\begin{equation}
\mathcal E(Z):=(\operatorname{Tr}_E Z)\otimes\tau_E,\qquad
\tau_E=I_E/d_E,\qquad
\widehat\omega=\omega_R\otimes\tau_E,\qquad
\widehat\sigma=\sigma_R\otimes\tau_E.
\label{eq:app-gw-expectation}
\end{equation}
This map is unital and preserves the trace.

Identify a matrix $Y$ with the Hilbert--Schmidt vector $\lvert Y\rangle\!\rangle$, whose inner product with $\lvert Z\rangle\!\rangle$ is $\operatorname{Tr}(Y^\dagger Z)$.
The relative modular operators and the isometry from $L_2(\mathcal N)$ to $L_2(\mathcal M)$ are defined as
\begin{align}
\Delta_{\mathcal M}(Y)&:=\sigma Y\omega^{-1},&
\Delta_{\mathcal N}(Y)&:=\widehat\sigma Y\widehat\omega^{-1},&
V_\omega\lvert Y\rangle\!\rangle
&:=\lvert Y\widehat\omega^{-1/2}\omega^{1/2}\rangle\!\rangle
\quad (Y\in\mathcal N).
\label{eq:app-gw-modular}
\end{align}
For a function $f$ on $(0,\infty)$, its quantum $f$-divergence is 
\begin{equation}
Q_f(\omega\Vert\sigma)
:=\langle\!\langle\omega^{1/2}\rvert
 f(\Delta_{\mathcal M})
 \lvert\omega^{1/2}\rangle\!\rangle,
\label{eq:app-gw-divergence}
\end{equation}
with the analogous definition using $\Delta_{\mathcal N}$ for
$Q_f(\widehat\omega\Vert\widehat\sigma)$. Finally, for any $\lambda>0$, define
\begin{equation}
w_\lambda
:=(\Delta_{\mathcal M}+\lambda)^{-1}\lvert\omega^{1/2}\rangle\!\rangle
-V_\omega(\Delta_{\mathcal N}+\lambda)^{-1}
 \lvert\widehat\omega^{1/2}\rangle\!\rangle.
\label{eq:app-gw-resolvent}
\end{equation}

We now state the two lemmas in \cite{GaoWilde2021Recoverability}.
\begin{lem}[Gao--Wilde~\cite{GaoWilde2021Recoverability}, Lemma~4.10, $t=0$]\label{lem:app-gw-410}
With the notation above, put
\begin{align}
v_0
&=\lvert\omega^{1/2}\rangle\!\rangle
-\Delta_{\mathcal M}^{1/2}V_\omega\Delta_{\mathcal N}^{-1/2}
 \lvert\widehat\omega^{1/2}\rangle\!\rangle \notag\\
&=\lvert\omega^{1/2}\rangle\!\rangle
-\lvert\sigma^{1/2}\widehat\sigma^{-1/2}
 \widehat\omega^{1/2}\rangle\!\rangle.
\label{eq:app-gw-vector}
\end{align}
Then
\begin{equation}
v_0=\frac{1}{\pi}\Delta_{\mathcal M}^{1/2}
\int_0^\infty\lambda^{-1/2}w_\lambda\,d\lambda,
\qquad
\norm{\omega-\mathcal R_\sigma(\widehat\omega)}_1
\le 2\norm{v_0}_2,
\label{eq:app-gw-lemma410}
\end{equation}
where the unrotated Petz map on $\mathcal N$ is
\begin{equation}
\mathcal R_\sigma(Y)
=\sigma^{1/2}\widehat\sigma^{-1/2}Y
 \widehat\sigma^{-1/2}\sigma^{1/2}.
\label{eq:app-gw-petz}
\end{equation}
\end{lem}

\begin{lem}[Gao--Wilde~\cite{GaoWilde2021Recoverability}, Lemma~4.11, $t=0$]\label{lem:app-gw-411}
Let $f:(0,\infty)\to\mathbb R$ be a regular operator anti-monotone
function, and let $d\nu$ be the positive measure in its integral
representation
\begin{equation}
f(x)=\alpha-\beta x+
\int_0^\infty\left(\frac{1}{\lambda+x}
-\frac{\lambda}{1+\lambda^2}\right)d\nu(\lambda),
\qquad \alpha\in\mathbb R,\quad \beta\ge0.
\label{eq:app-gw-representation}
\end{equation}
Here regularity means that $d\nu$ is absolutely continuous with respect
to Lebesgue measure and that, for every $0<a<b<\infty$, there is a finite
$c(a,b)>0$ such that $d\lambda\le c(a,b)\,d\nu(\lambda)$ on $(a,b)$.
Then for any $0<a<b<\infty$, 
\begin{equation}
\norm{v_0}_2\le\frac{1}{\pi}\left(
4\sqrt{a q}+
\sqrt{c(a,b)\log(b/a)
 \left[Q_f(\omega\Vert\sigma)
       -Q_f(\widehat\omega\Vert\widehat\sigma)\right]}
+\frac{4}{\sqrt b}\right)
\label{eq:app-gw-lemma411}
\end{equation}
where $q:=Q_{x^{-1}}(\omega\Vert\sigma)
=\operatorname{Tr}(\omega^2\sigma^{-1})$.
\end{lem}

% We use this vector estimate, rather than only the trace-distance
% consequence in \eqref{eq:app-gw-lemma410}.

\begin{proof}[Proof of Lemma~\ref{lem:main}]

For $f(x)=-\log x$, the representation
\begin{equation}
-\log x=\int_0^\infty
\left(\frac{1}{\lambda+x}-\frac{\lambda}{1+\lambda^2}\right)d\lambda
\label{eq:app-gw-log}
\end{equation}
shows that $c(a,b)=1$.
Since left and right multiplication commute, functional calculus gives
$Q_{-\log}(\omega\Vert\sigma)=D(\omega\Vert\sigma)$ and
$Q_{x^{-1}}(\omega\Vert\sigma)=\operatorname{Tr}(\omega^2\sigma^{-1})$.
The common factor $\tau_E$ cancels in both
\begin{equation}
D(\widehat\omega\Vert\widehat\sigma)=D(\omega_R\Vert\sigma_R),
\qquad
\widehat\sigma^{-1/2}\widehat\omega^{1/2}
=(\sigma_R^{-1/2}\omega_R^{1/2})\otimes I_E.
\label{eq:app-gw-cancellation}
\end{equation}
Thus, with $\eta=D(\omega\Vert\sigma)-D(\omega_R\Vert\sigma_R)$,
the two lemmas yield
\begin{equation}
\norm{\omega^{1/2}-\sigma^{1/2}
(\sigma_R^{-1/2}\omega_R^{1/2}\otimes I_E)}_2
\le\frac{1}{\pi}\left(
4\sqrt{a q}+\sqrt{\eta\log(b/a)}+\frac{4}{\sqrt b}\right),
\qquad 0<a<b.
\label{eq:app-gw-partial-trace}
\end{equation}
No dimension factor arises from the translation to a partial trace.

\emph{Specialization to the tripartite state.}
First assume $\rho=\rho_{ABC}>0$. In
\eqref{eq:app-gw-partial-trace}, choose
\begin{equation}
R=AB,\qquad E=C,\qquad \omega=\rho,\qquad
\sigma=\tau_A\otimes\rho_{BC},\qquad
\tau_A=I_A/d_A,\qquad \sigma_R=\tau_A\otimes\rho_B.
\label{eq:app-gw-choice}
\end{equation}
The relative-entropy difference becomes
\begin{equation}
\eta
=D(\rho\Vert\tau_A\otimes\rho_{BC})
 -D(\rho_{AB}\Vert\tau_A\otimes\rho_B) 
 =I(A:C\mid B)_\rho.
\label{eq:app-gw-cmi}
\end{equation}
Define the amplitude and its error by
\begin{equation}
X_\rho=\rho_{AB}^{1/2}\rho_B^{-1/2}\rho_{BC}^{1/2},
\qquad \varepsilon=\norm{\rho^{1/2}-X_\rho}_2.
\label{eq:app-gw-target}
\end{equation}
The second term on the l.h.s. of \eqref{eq:app-gw-partial-trace} is, after substitution,
\begin{align}
(\tau_A^{1/2}\otimes\rho_{BC}^{1/2})
(\tau_A^{-1/2}\otimes\rho_B^{-1/2}\otimes I_C)
(\rho_{AB}^{1/2}\otimes I_C)
&=\rho_{BC}^{1/2}\rho_B^{-1/2}\rho_{AB}^{1/2}
=X_\rho^\dagger.
\label{eq:app-gw-order}
\end{align}
Since $\rho^{1/2}$ is self-adjoint, $\norm{\rho^{1/2}-X_\rho^\dagger}_2=\varepsilon$. Consequently,
\begin{equation}
\varepsilon\le\frac{1}{\pi}
\left(4\sqrt{a q}+\sqrt{\eta\log(b/a)}+\frac{4}{\sqrt b}\right),
\qquad
q=d_A\operatorname{Tr}\!\left[\rho^2(I_A\otimes\rho_{BC}^{-1})\right],
\qquad 0<a<b.
\label{eq:app-gw-specialized}
\end{equation}

\emph{Bounding the relative second moment $q$.}
For every positive semidefinite $M_{AR}$, with
$M_R=\operatorname{Tr}_A M_{AR}$, one has
\begin{equation}
M_{AR}\le d_A I_A\otimes M_R,
\label{eq:app-gw-domination}
\end{equation}
due to the pinching inequality.
% Indeed, write $\lvert\psi\rangle=\sum_i\lvert i\rangle\lvert\psi_i\rangle$
% in an orthonormal basis of $A$, and put
% $M_{ii}=(\langle i\rvert\otimes I_R)M_{AR}(\lvert i\rangle\otimes I_R)$.
% The vector Cauchy--Schwarz inequality and
% $0\le M_{ii}\le\sum_jM_{jj}=M_R$ give
% \begin{align}
% \langle\psi\rvert M_{AR}\lvert\psi\rangle
% &=\left\|\sum_i M_{AR}^{1/2}
%        (\lvert i\rangle\lvert\psi_i\rangle)\right\|^2
%  \le d_A\sum_i\langle\psi_i\rvert M_{ii}\lvert\psi_i\rangle
%  \le d_A\langle\psi\rvert I_A\otimes M_R\lvert\psi\rangle.
% \label{eq:app-gw-domination-proof}
% \end{align}
Apply \eqref{eq:app-gw-domination} once, with $M_{AR}=\rho$ and $R=BC$.
Inversion reverses the positive-definite order, and congruence by $\rho$
preserves it, so
\begin{align}
\rho&\le d_A I_A\otimes\rho_{BC},
& I_A\otimes\rho_{BC}^{-1}&\le d_A\rho^{-1}, \notag\\
q&=d_A\operatorname{Tr}\!\left[
 \rho(I_A\otimes\rho_{BC}^{-1})\rho\right]
 \le d_A^2\operatorname{Tr}\rho=d_A^2.
\label{eq:app-gw-moment}
\end{align}
Substitution into \eqref{eq:app-gw-specialized} leaves the cutoffs free:
\begin{equation}
\varepsilon\le\frac{1}{\pi}
\left(4d_A\sqrt a+\sqrt{\eta\log(b/a)}+\frac{4}{\sqrt b}\right),
\qquad 0<a<b.
\label{eq:app-gw-free}
\end{equation}

\emph{Choosing $a$ and $b$.}
Interchanging $A$ and $C$ leaves $\eta$ unchanged, we can let  $d_*=\min\{d_A,d_C\}$, and
\begin{equation}
\varepsilon\le\frac{1}{\pi}
\left(4d_*\sqrt a+\sqrt{\eta\log(b/a)}+\frac{4}{\sqrt b}\right),
\qquad 0<a<b.
\label{eq:app-gw-symmetric}
\end{equation}
For $0<\eta\le1$, choose
\begin{equation}
a=\frac{\eta}{16d_*^2},\qquad b=\frac{16}{\eta}.
\label{eq:app-gw-cutoffs}
\end{equation}
Then $b/a=256d_*^2/\eta^2>1$,
$4d_*\sqrt a=4/\sqrt b=\sqrt\eta$, and
$\log(b/a)=2\log(16d_*/\eta)$. Hence
\begin{equation}
\varepsilon\le\frac{\sqrt\eta}{\pi}
\left(2+\sqrt{2\log\frac{16d_*}{\eta}}\right),
\label{eq:app-gw-final}
\end{equation}
which is \eqref{eq:main}. 

\end{proof}

\subsection{Exact quantum Markov chains}\label{app: exact}

\begin{thm}\label{thm:exact}
    Given a state $\rho_{AB}$ and a channel $\mathcal{N}_{B\to C}$. Let $\sigma_{AC} := \mathcal{N}_{B\to C}(\rho_{AB})$. There exists a recovery channel $\mathcal{R}_{C\to B}$ such that $\rho_{AB} = \mathcal{R}_{C\to B}(\sigma_{AC})$ if and only if there exists an isometry $W_{CC'\to BB'}$ such that $\left|\psi_\rho\right\rangle_{A A^{\prime} B B^{\prime}}=W_{C C^{\prime} \rightarrow B B^{\prime}}\left|\psi_\sigma\right\rangle_{A A^{\prime} C C^{\prime}}$ 
\end{thm}
\noindent As a comment, Theorem \ref{thm:exact} does not imply mixed-state LR if we do not assume that there exists already a one-way channel $\mathcal{N}_{B\to C}$.

First, we prove the only-if direction of Theorem~\ref{thm:exact}.

\begin{prop}\label{prop:app exact forward}
Let $\rho_{ABC}$ be a quantum Markov chain that satisfies $I(A:C|B) = 0$. Let $|\psi_{\rho_{ABC}}\rangle_{AA'BB'CC'}$ be the canonical purification of $\rho_{ABC}$ and $|\psi_{\rho_{AB}}\rangle_{AA'BB'}$ be the canonical purification of $\rho_{AB}$, then there exists a unitary $U_{BB'CC'}$ such that $|\psi_{\rho_{ABC}}\rangle_{AA'BB'CC'} = U_{BB'CC'}(|\psi_{\rho_{AB}}\rangle_{AA'BB'}\otimes |0\rangle_{CC'})$
\end{prop}
\begin{proof}
We use the structure theorem of the Markov chain
\begin{equation}
    \rho_{ABC} = \sum_{i} p_i \rho_{AB^{i}_L} \otimes \rho_{B^i_R C}
\end{equation}
where we have decomposed the Hilbert space of $B$ into $\mathcal{H}_B = \oplus_{i} (\mathcal{H}_{B^{i}_L} \otimes \mathcal{H}_{B^{i}_R})$. In this decomposition each subspace $\mathcal{H}_{B^{i}_{L/R}}$ is orthogonal to the other subspaces, i.e. $\rho_{AB^{i}_{L/R}} \rho_{AB^{j}_{L/R}} = 0 ,\; (i\neq j).$
This allows us to take the square root directly,
\begin{equation}
    \sqrt{\rho_{ABC}} = \sum_{i} \sqrt{p_i} \sqrt{\rho_{AB^{i}_L}} \otimes \sqrt{\rho_{B^i_R C}}
\end{equation}
On the other hand, we can trace out $C$ and obtain
\begin{equation}
    \sqrt{\rho_{AB}} = \sum_{i} \sqrt{p_i} \sqrt{\rho_{AB^{i}_L}} \otimes \sqrt{\rho_{B^i_R}}
\end{equation}
Therefore,
\begin{align}
    |\psi_{\rho_{ABC}}\rangle_{AA'BB'CC'} &=\sum_{i} \sqrt{p_i} \sqrt{\rho_{AB^i_L}}|\Omega\rangle_{AA'B^{i}_LB^{'i}_L} \otimes \sqrt{\rho_{B^i_R C}} |\Omega\rangle_{B^{i}_RB^{'i}_R CC'} \nonumber\\
    |\psi_{\rho_{AB}}\rangle_{AA'BB'} &=\sum_{i} \sqrt{p_i} \sqrt{\rho_{AB^i_L}}|\Omega\rangle_{AA'B^{i}_LB^{'i}_L} \otimes \sqrt{\rho_{B^i_R}} |\Omega\rangle_{B^{i}_RB^{'i}_R}
\end{align}
Now we can trace out the rest of the system to obtain the reduced density matrix on $AA'$, and we find
\begin{equation}
    \tr_{BB'CC'}(|\psi_{\rho_{ABC}}\rangle\langle |\psi_{\rho_{ABC}}|) = \tr_{BB'}(|\psi_{\rho_{AB}}\rangle\langle |\psi_{\rho_{AB}}|) = \sum_{i} p_i \tr_{B^i_L} \sqrt{\rho_{AB^i_L}}(|I\rangle_{AA'B^{i}_LB^{'i}_L} \langle I|)\sqrt{\rho_{AB^i_L}}.
\end{equation}
This means that $|\psi_{\rho_{ABC}}\rangle$ and $|\psi_{\rho_{AB}}\rangle$ are purifications of the same reduced density matrix; therefore, they must be related by an isometry $W_{BB'\rightarrow BB'CC'}$,
\begin{equation}
    |\psi_{\rho_{ABC}}\rangle_{AA'BB'CC'} = W_{BB'\rightarrow BB'CC'}|\psi_{\rho_{AB}}\rangle_{AA'BB'}
\end{equation}
This isometry can be implemented by first adding ancilla $|0\rangle_{CC'}$ and then acting on a unitary
\begin{equation}
    |\psi_{\rho_{ABC}}\rangle_{AA'BB'CC'} = U_{BB'CC'}(|\psi_{\rho_{AB}}\rangle_{AA'BB'}\otimes |0\rangle_{CC'})
\end{equation}
which completes the proof.
\end{proof}

A useful lemma is the following. A contraction is a linear operator that does not increase the norm of any vector.

\begin{lem}\label{app:lem contraction to isometry}
If two pure states are approximately connected by a contraction acting on subsystems
\begin{align}
    \abs{\langle \psi_{AB} | \mathcal{C}_{C\to B}|\phi_{AC}\rangle}\geq 1-\delta,
\end{align}
then, there exists an isometry $W_{C\to B}$, assuming $d_B\geq d_C$, such that
\begin{align}
    \abs{\langle \psi_{AB} | W_{C\to B}|\phi_{AC}\rangle}\geq 1-\delta
\end{align}
When $\delta = 0$, the isometry is exact.
\end{lem}
\begin{proof}
    Observe that
    \begin{align}
        \abs{\langle \psi_{AB}| I_{A} \otimes \mathcal{C}_{C\to B} |\phi_{AC}\rangle } \leq \norm{\mathcal{C}_{C\to B}} \norm{\tr_{A}|\phi_{AC}\rangle\langle\psi_{AB}|}_1 \leq F(\tr_{B}|\psi_{AB}\rangle\langle \psi_{AB}| ,\tr_{C}|\phi_{AC}\rangle\langle \phi_{AC}|),
    \end{align}
    where in the second inequality we used that $\norm{\mathcal{C}_{C\to B}}\leq 1$ is a contraction. To see that the 1-norm term gives the fidelity, we decompose $\tr_{B}|\psi_{AB}\rangle\langle \psi_{AB}| = XX^\dagger$ and $\tr_{C}|\phi_{AC}\rangle\langle \phi_{AC}| = YY^\dagger$. Since  $\tr_{A}|\phi_{AC}\rangle\langle\psi_{AB}| = (X^\dagger Y)^T$, we have $F(XX^\dagger,YY^\dagger) = \norm{\sqrt{XX^\dagger}\sqrt{YY^\dagger} }_1 = \norm{X^\dagger Y}_1 = \norm{\tr_{A}|\psi_{AC}\rangle\langle\psi_{AB}|}_1$.
    Therefore, by Uhlmann theorem, there exists an isometry $W_{C\to B}$ such that
    \begin{align}
    \abs{\langle \psi_{AB} | W_{C\to B}|\phi_{AC}\rangle}\geq 1-\delta
    \end{align}
    provided $d_B\geq d_C$. When $\delta = 0$, as the overlap is upper bounded by $1$,
    \begin{align}
    \abs{\langle \psi_{AB} | W_{C\to B}|\phi_{AC}\rangle}=1.
    \end{align}
\end{proof}

\begin{cor}\label{cor: exact forward}
Consider any quantum channel $\mathcal{N}_{B\rightarrow C}$ on a bipartite mixed state $\rho_{AB}$, which gives $\sigma_{AC} = \mathcal{N}(\rho_{AB})$, such that there exists a reverse channel $\mathcal{R}_{C\rightarrow B}$, $\rho_{AB} = \mathcal{R}(\rho_{AC})$. The canonical purifications are denoted as $|\psi_{\rho}\rangle_{AA'BB'}$ and $|\psi_{\sigma}\rangle_{AA'CC'}$. Then there exists an isometry $W_{CC'\rightarrow BB'}$ such that $W_{CC'\rightarrow BB'}|\psi_{\sigma}\rangle_{AA'CC'} = |\psi_{\rho}\rangle_{AA'BB'}$.
\end{cor}
\begin{proof}
    The reversibility condition implies that $I(A:B)_{\rho} = I(A:C)_{\sigma}$. Now consider a Steinspring dilation $V_{B\rightarrow CE}$ of the channel $\mathcal{N}$, such that
    \begin{equation}
        \sigma_{AC} = \tr_E \gamma_{ACE}
    \end{equation}
    where
    \begin{equation}
        \gamma_{ACE} = V_{B\rightarrow CE} \rho_{AB} V^{\dagger}_{B\rightarrow CE}
    \end{equation}
    and $E$ is the environment.
Then $I(A:B)_{\rho} = I(A:CE)_{\gamma}$. By reversibility we also have $I(A:B)_{\rho} = I(A:C)_{\sigma} = I({A:C)_{\gamma}}$, which gives rise to
\begin{equation}
    I(A:E|C)_{\gamma} = 0
\end{equation}
It follows from the Proposition~\ref{prop:app exact forward} that the canonical purification $|\psi_{\gamma}\rangle_{AA'CC'EE'}$ of $\gamma$ is related to the canonical purification $|\psi_{\sigma}\rangle_{AA'CC'}$ by an isometry $\tilde{W}$,
\begin{equation}
    |\psi_{\gamma}\rangle_{AA'CC'EE'} = \tilde{W}_{CC'\rightarrow CC'EE'} |\psi_{\sigma}\rangle_{AA'CC'}
\end{equation}
Since we also have
    \begin{equation}
        |\psi_{\gamma}\rangle_{AA'CC'EE'} = (V_{B\rightarrow CE} \otimes V_{B'\rightarrow C'E'})|\psi_{\rho}\rangle_{AABB'},
    \end{equation}
the two states on the right-hand side are related by a contraction
\begin{equation}
    \mathcal{C}_{CC'\rightarrow BB'}|\psi_{\sigma}\rangle_{AA'CC'} = |\psi_{\rho}\rangle_{AA'BB'}
\end{equation}
where $\mathcal{C}_{CC'\rightarrow BB'} :=(V^{\dagger}_{B\rightarrow CE} \otimes V^{\dagger}_{B'\rightarrow C'E'})\tilde{W}_{CC'\rightarrow CC'EE'}$ and the adjoint of the isometry is a contraction. Applying Lemma \ref{app:lem contraction to isometry} completes the proof.
\end{proof}

Next, we prove the if-direction of Theorem~\ref{thm:exact}

\begin{lem}\label{lem:Wilde-Holevo}
Let $\rho,\sigma$ be density operators such that $\operatorname{supp}(\rho)\subseteq \operatorname{supp}(\sigma)$, and let $\mathcal N$ be a quantum channel. Define the Holevo fidelity
\begin{equation}
    F_H(\rho,\sigma):=\operatorname{Tr}(\sqrt{\rho}\sqrt{\sigma}).
\end{equation}
We have
\begin{equation}
    F_H(\mathcal N(\rho),\mathcal N(\sigma)) = F_H(\rho,\sigma)\quad \Rightarrow \quad \mathcal R_{\sigma,\mathcal N}(\mathcal N(\rho))=\rho,
\end{equation}
where $\mathcal R_{\sigma,\mathcal N}$ is the Petz recovery map
\begin{equation}
    \mathcal R_{\sigma,\mathcal N}(X)
    :=
    \sigma^{1/2}\,
    \mathcal N^*\!\left(
        \mathcal N(\sigma)^{-1/2}
        X
        \mathcal N(\sigma)^{-1/2}
    \right)
    \sigma^{1/2},
\end{equation}
\end{lem}

\begin{proof}
This is an immediate consequence of Wilde's recovery theorem in \cite{wilde_recoverability_2018}: 
\begin{equation}
    F_H(\mathcal N(\rho),\mathcal N(\sigma)) - F_H(\rho,\sigma)
    \ge c_\sigma \,
    \bigl\|\mathcal R_{\sigma,\mathcal N}(\mathcal N(\rho))-\rho\bigr\|_1^3,
\end{equation}
where $c_\sigma>0$ is an explicit constant depending on $\sigma$.
\end{proof}

\begin{prop}\label{prop:reverse-exact}
Let $\rho_{AB}$ be a bipartite state, and let $\sigma_{AC} = \mathcal N_{B\to C}(\rho_{AB})$
for some channel $\mathcal N_{B\to C}$. Let $|\psi_\rho\rangle_{AA'BB'}$ and $|\psi_\sigma\rangle_{AA'CC'}$ denote the canonical purifications of $\rho_{AB}$ and $\sigma_{AC}$ respectively. If the reduced states on $AA'$ are equal, i.e.
\begin{equation}
    \operatorname{Tr}_{BB'} |\psi_\rho\rangle\langle\psi_\rho|
    =
    \operatorname{Tr}_{CC'} |\psi_\sigma\rangle\langle\psi_\sigma|,
\end{equation}
then there exists a recovery channel $\mathcal R_{C\to B}$ such that
\begin{equation}
     \mathcal R_{C\to B}(\sigma_{AC})=\rho_{AB}.
\end{equation}
\end{prop}

\begin{proof}
For any unitary $U_A$ on $A$, define
\begin{equation}
    \rho_{AB}^U := U_A \rho_{AB} U_A^\dagger,\quad 
    \sigma_{AC}^U := U_A \sigma_{AC} U_A^\dagger
\end{equation}
Then
\begin{align}
    F_H(\rho_{AB}^U,\rho_{AB})
    &= \operatorname{Tr}\!\left(U_A \sqrt{\rho_{AB}} U_A^\dagger \sqrt{\rho_{AB}}\right) = \langle \psi_\rho |\, U_A \otimes U_{A'}^* \,| \psi_\rho\rangle,\\
    F_H(\sigma_{AC}^U,\sigma_{AC})
    &= \operatorname{Tr}\!\left(U_A \sqrt{\sigma_{AC}} U_A^\dagger \sqrt{\sigma_{AC}}\right) = \langle \psi_\sigma |\, U_A \otimes U_{A'}^* \,| \psi_\sigma\rangle
\end{align}
Since $|\psi_\rho\rangle$ and $|\psi_\sigma\rangle$ have the same reduced state on $AA'$, it follows that for every unitary $U_A$,
\begin{equation}
    F_H(\sigma_{AC}^U,\sigma_{AC})
    =
    F_H(\rho_{AB}^U,\rho_{AB}).
\end{equation}
Because $\sigma_{AC}=\mathcal N_{B\to C}(\rho_{AB})$ and $\sigma_{AC}^U=\mathcal N_{B\to C}(\rho_{AB}^U)$, this becomes
\begin{equation}
    F_H(\mathcal N(\rho_{AB}^U),\mathcal N(\rho_{AB}))
    =
    F_H(\rho_{AB}^U,\rho_{AB}).
\end{equation}
Now applying Lemma~\ref{lem:Wilde-Holevo} with $\rho=\rho_{AB}^U$ and $\sigma=\rho_{AB}$, we have
\begin{equation}
    \mathcal R_{\rho_{AB},\mathcal N}(\sigma_{AC}^U)=\rho_{AB}^U
    \qquad \forall\, U_A.
\end{equation}
Next, we average over Haar-random $U_A$ which results in a completely depolarizing channel on $A$:
\begin{equation}
    \int dU_A \, \rho_{AB}^U
    =
    \frac{I_A}{d_A}\otimes \rho_B, \quad \int dU_A \, \sigma_{AC}^U
    =
    \frac{I_A}{d_A}\otimes \sigma_C.
\end{equation}
By linearity of $\mathcal R_{\rho_{AB},\mathcal N}$,
\begin{equation}
    \mathcal R_{\rho_{AB},\mathcal N}\!\left(\frac{I_A}{d_A}\otimes \sigma_C\right)
    =
    \frac{I_A}{d_A}\otimes \rho_B.
\end{equation}
Then, using monotonicity of quantum relative entropy, we observe
\begin{align}
    D\!\left(\sigma_{AC}\,\|\,\frac{I_A}{d_A}\otimes \sigma_C\right)
    &\ge
    D\!\left(
        \mathcal R_{\rho_{AB},\mathcal N}(\sigma_{AC})
        \,\|\, 
        \mathcal R_{\rho_{AB},\mathcal N}\!\left(\frac{I_A}{d_A}\otimes \sigma_C\right)
    \right)=
    D\!\left(\rho_{AB}\,\|\,\frac{I_A}{d_A}\otimes \rho_B\right),\\
    D\!\left(\rho_{AB}\,\|\,\frac{I_A}{d_A}\otimes \rho_B\right)
    &\ge
    D\!\left(
        \mathcal N_{B\to C}(\rho_{AB})
        \,\|\, 
        \mathcal N_{B\to C}\!\left(\frac{I_A}{d_A}\otimes \rho_B\right)
    \right) =
    D\!\left(\sigma_{AC}\,\|\,\frac{I_A}{d_A}\otimes \sigma_C\right).
\end{align}
Therefore,
\begin{equation}
    D\!\left(\rho_{AB}\,\|\,\frac{I_A}{d_A}\otimes \rho_B\right)
    =
    D\!\left(\sigma_{AC}\,\|\,\frac{I_A}{d_A}\otimes \sigma_C\right),
\end{equation}
which leads to
\begin{align}
    I(A:B)_\rho = D(\rho_{AB}\,\|\,\rho_A\otimes \rho_B) = D(\sigma_{AC}\,\|\,\sigma_A\otimes \sigma_C) = I(A:C)_\sigma 
\end{align}
By Petz's sufficiency theorem, there exists a recovery channel $\mathcal R_{C\to B}$ such that
\begin{align}
    \mathcal R_{C\to B}(\sigma_{AC})=\rho_{AB}.
\end{align}
\end{proof}

\subsection{approximate classical and quantum stabilizer Markov chains }\label{app:approx classical}

Consider a classical distribution $p_{ABC}(a,b,c)$ and the canonical purification is given by 
\begin{equation}
    |\psi_p\rangle = \sum_{a,b,c} \sqrt{p_{ABC}(a,b,c)}|abc\rangle_{ABC}\otimes|abc\rangle_{A'B'C'}
\end{equation}
Here are some facts. First, $|\langle \psi_p|\psi_q\rangle| = F(p,q)$, where $F(p,q) := \tr \sqrt{p}\sqrt{q}$ is the Uhlmann fidelity, which is equal to the Holevo fidelity for classical distributions.
This ensures that two close distributions have close canonical purifications. Another fact is that $F(p_{ABC}, p_{AB}p_{C|B})\geq e^{-I(A:C|B)/2}$.
We can now state the following Proposition:
\begin{prop}\label{prop:app approx forward classical}
    Let $p_{ABC}$ be a classical distribution with $I(A:C|B) \leq \epsilon$. Let $|\psi_{p_{ABC}}\rangle_{AA'BB'CC'}$ be the canonical purification. Also, let $|\psi_{p_{AB}}\rangle_{AA'BB'}$ be the canonical purification of $p_{AB}$. Then, there exists a unitary $U_{BB'CC'}$ such that
    \begin{equation}
        |\langle\psi_{p_{ABC}}| U_{BB'CC'} (|\psi_{p_{AB}}\rangle\otimes |0\rangle_{CC'}) |\geq e^{-\epsilon/2}
    \end{equation}
\end{prop}
\begin{proof}
    Consider the distribution $r_{ABC} = p_{AB}p_{C|B}$ and its canonical purification $|\psi_{r_{ABC}}\rangle$. Then we have
    \begin{equation}
        |\langle \psi_{p_{ABC}}|\psi_{r_{ABC}}\rangle| = F(p_{ABC},r_{ABC}) \geq e^{-\epsilon/2}
    \end{equation}
    By data processing inequality, this implies that the reduced density matrix on $AA'$ are similar, i.e.,
    \begin{equation}
        F(\tr_{BB'CC'}(|\psi_{p_{ABC}}\rangle \langle \psi_{p_{ABC}}|), \tr_{BB'CC'}(|\psi_{r_{ABC}}\rangle \langle \psi_{r_{ABC}}|))\geq e^{-\epsilon/2}
    \end{equation}
    From Proposition \ref{prop:app exact forward} we know that $\tr_{BB'CC'}(|\psi_{r_{ABC}}\rangle \langle \psi_{r_{ABC}}|)$ is also the reduced density matrix of $|\psi_{p_{AB}}\rangle$ on $AA'$. Therefore, by the Uhlmann theorem, there exists a unitary such that
    \begin{equation}
        |\langle\psi_{p_{ABC}}| U_{BB'CC'} (|\psi_{p_{AB}}\rangle\otimes |0\rangle_{CC'}) | = F(\tr_{BB'CC'}(|\psi_{p_{ABC}}\rangle \langle \psi_{p_{ABC}}|), \tr_{BB'}(|\psi_{r_{AB}}\rangle \langle \psi_{r_{AB}}|)) \geq e^{-\epsilon/2}.
    \end{equation}
\end{proof}

Next, consider a quantum stabilizer state $\rho_{ABC}$ that satisfies
\begin{align}
    [\rho_{AB},\rho_{BC}] = 0.
\end{align}

\begin{prop}\label{prop:app approx forward commuting}
    Let $\rho_{ABC}$ be a quantum stabilizer state with $I(A:C|B) \leq  \epsilon$, then there exists a unitary $U_{BB'CC'}$ such that
    \begin{equation}
        |\langle \psi_{\rho_{ABC}}|U_{BB'CC'}(|\psi_{\rho_{AB}}\rangle\otimes |0\rangle_{CC'})|\geq e^{-\epsilon/2}
    \end{equation}
\end{prop}
\begin{proof}
    Let $|\psi_{\sigma_{ABC}}\rangle$ be the canonical purification  of $\sigma_{ABC}:=\rho^{1/2}_{BC}\rho^{-1/2}_{B}\rho_{AB}\rho^{-1/2}_{B}\rho^{1/2}_{BC}$. Since $[\rho_{BC},\rho_{AB}]=0$, we also have $[\rho_{BC},\rho_B]=0$ and $[\rho_{AB},\rho_B]=0$, and $\tr_{C}\sigma_{ABC} = \tr_C \rho_{AB}\rho^{-1}_{B}\rho_{BC} =  \rho_{AB}\rho^{-1}_{B} \tr_C\rho_{BC} = \rho_{AB}$. Therefore $\sigma_{ABC}$ is an exact quantum Markov chain. It follows from the monotonicity of Renyi relative entropy that 
    \begin{equation}
        \langle \psi_{\rho_{ABC}}|\psi_{\sigma_{ABC}}\rangle = F_H(\rho,\sigma)\geq e^{-S(\rho||\sigma)/2} \geq e^{-\epsilon/2}, 
    \end{equation}
    where we used $S(\rho_{ABC}||\sigma_{ABC}) = I(A:C|B)$. Now, we can repeat the reasoning in Proposition \ref{prop:app approx forward classical}.
\end{proof}

\begin{cor}
    Let $\rho_{AB}$ be a classical distribution or a quantum commuting mixed state, and let $\sigma_{AC} = \mathcal N_{B\to C}(\rho_{AB})$
for some channel $\mathcal N_{B\to C}$. If $\epsilon := I(A:B)_\rho - I(A:C)_\sigma$, then there exists an isometry $W_{CC'\to BB'}$, such that $|\langle\psi_\rho|W_{CC'\to BB'}  |\psi_\sigma\rangle| \geq e^{-\epsilon/2}$
\end{cor}

\begin{proof}
The same as Corollary \ref{cor: exact forward}.
\end{proof}

\subsection{approximate quantum non-commuting Markov chains }\label{app:approx quantum}

We prove Theorem~\ref{thm:main rotated petz} using Junge-LaRacuent's \cite{junge_multivariate_2022} Corollary 6.1 

\begin{lem}[Corollary 6.1 in \cite{junge_multivariate_2022}]\label{thm: app universal recovery}
    Let $\rho,\sigma$ be density matrices and $\mathcal{N}$ a channel. Denote $\hat \rho:= \mathcal{N}(\rho), \hat \sigma:= \mathcal{N}(\sigma)$. We have
    \begin{align}\label{eq: f2}
        -\log f_2\left( \rho^{1/2}, \tilde{\mathcal{R}}_{\sigma,\mathcal{N}}(\hat \rho^{1/2})\right) \leq \frac{1}{4}(D(\rho||\sigma) - D(\hat \rho||\hat \sigma)),
    \end{align}
    where $f_2(\rho,\sigma):= ||\sqrt{\rho}\sqrt{\sigma}||_2$ and $D(\rho||\sigma)  = \tr(\rho(\log\rho - \log \sigma))$. The universal recovery map $\tilde{\mathcal{R}}_{\sigma,\mathcal{N}}$ is given by
    \begin{align}
        \tilde{\mathcal{R}}_{\sigma,\mathcal{N}}(X):= \int_{\mathbb{R}} \mathcal{R}^t_{\sigma,\mathcal{N}}(X) \beta_0(t) \ud t, 
    \end{align}
    where $\beta_0(t):= \pi/(2(\cosh(\pi t)+1))$ and
    \begin{align}
        \mathcal{R}^t_{\sigma,\mathcal{N}}(X):= \sigma^{(1-\ii t)/4} \mathcal{N}^\dagger\left(\hat{\sigma}^{-(1-\ii t)/4} X \hat{\sigma}^{-(1+\ii t)/4}\right)\sigma^{(1+\ii t)/4}
    \end{align}
    Moreover, the map $\tilde{\mathcal{R}}_{\sigma,\mathcal{N}}$ is a contraction, i.e. $||\tilde{\mathcal{R}}_{\sigma,\mathcal{N}}(X)||_2 \leq ||X||_2$.
\end{lem}
This Lemma uses the $p=2$ case of the $p$-norm results in \cite{junge_multivariate_2022}.

\begin{proof}[Proof of Theorem~\ref{thm:main rotated petz}]
 Apply Lemma~\ref{thm: app universal recovery} to $\mathcal{N} = I_A \otimes \mathcal{N}_{B\to C}$, $\rho = \rho_{AB}$, $\sigma = \rho_A \otimes \rho_B$. First, we have
    \begin{align}
        D(\rho_{AB}|| \rho_A \otimes \rho_B) - D(\sigma_{AC}|| \sigma_A \otimes \sigma_B) = \epsilon.
    \end{align}
    Second, the universal recovery map induces a contraction as follows
    \begin{align}
        \mathcal{C}_{CC'\to BB'}|X_{AC}\rangle:= |\tilde{\mathcal{R}}_{\rho_A \otimes \rho_B, I_A \otimes \mathcal{N}_{B\to C}}(X_{AC}) \rangle= |(I_A\otimes \tilde{\mathcal{C}})(X_{AC})\rangle
    \end{align}
    where $\tilde{\mathcal{C}}(Y_{C}) := \int_{\mathbb{R}} \tilde{\mathcal{C}}^t(Y_{C}) \beta_0(t)\ud t$ and 
    \begin{align}
       \tilde{\mathcal{C}}^t(Y_{C})  := \rho_{B}^{(1-\ii t)/4} \mathcal{N}^\dagger\left(\sigma_C^{-(1-\ii t)/4} Y_C \sigma_C^{-(1+\ii t)/4}\right) \rho_{B}^{(1+\ii t)/4}. 
    \end{align}
    Then, \eqref{eq: f2} leads to
    \begin{align}
        \abs{\langle \psi_\rho| I_{A A^{\prime}} \otimes \mathcal{C}_{CC'\to BB'} |\psi_\sigma\rangle } = f_2(\rho_{AB}^{1/2}, \tilde{\mathcal{R}}(\sigma_{AC}^{1/2}))^2 \geq e^{-\epsilon/2}
    \end{align}
    Applying Lemma \ref{app:lem contraction to isometry} completes the proof.
\end{proof}

\section{Antisymmetric-state example: small CMI without closeness to a Markov chain} \label{app:antisymmetric state}

A standard obstruction to the naive slogan ``small conditional mutual information implies closeness to an exact quantum Markov chain'' is provided by the antisymmetric state. For this family, one can construct extensions with \(I(A:B|E)\to 0\) as \(d\to\infty\) where $d$ is the local Hilbert space dimension, while the state remains a constant trace distance away from every exact Markov chain. At the same time, the canonical purifications of \(\rho_{ABE}\) and \(\rho_{AE}\) have reduced density matrices on \(AA'\) that become asymptotically identical. This shows that closeness of the \(AA'\) reduced density matrices of canonical purifications is a much weaker property than closeness to the set of exact Markov chains.

Let \(A,B \simeq \mathbb{C}^d\), and for each \(k\in \{2,\dots,d\}\), let $P_k$
denote the projector onto the totally antisymmetric subspace
\begin{align}
    \bigwedge\nolimits^k(\mathbb{C}^d)\subset (\mathbb{C}^d)^{\otimes k}.
\end{align}
We write $d_k := \dim \bigwedge\nolimits^k(\mathbb{C}^d)=\binom{d}{k}$. We regard the first tensor factor as \(A\), the second as \(B\), and the remaining \(k-2\) factors as \(E\). Define
\begin{align}
    \rho_{ABE}^{(k)} := \frac{P_k}{d_k}.
\end{align}
Then, the reduced density matrices are given by
\begin{align}
    \rho_{AE}^{(k)} = \operatorname{Tr}_B \rho_{ABE}^{(k)} = \frac{P_{k-1}}{d_{k-1}}, \quad \rho_{AB}^{(k)}=\frac{2}{d(d-1)}P_2,
\end{align}
since the $n$-particle reduction of the maximally mixed state on \(\bigwedge^k(\mathbb{C}^d)\) is the maximally mixed state on \(\bigwedge^{k-n}(\mathbb{C}^d)\). The projector $P_2$ can also be written using swap operator $F$ through $P_2 = (I-F)/2$.

For the above extension, one has the exact formula
\begin{align}
    I(A:B|E)_{\rho^{(k)}} &= S(AE)+S(BE) - S(E) - S(ABE) \nonumber\\
    & = 2 \log d_{k-1} - \log(d_{k-2}) - \log(d_k)\nonumber\\
&=
\log\!\left(\frac{k}{k-1}\frac{d-k+2}{d-k+1}\right),
\end{align}
where we used $S(P_k/d_k) = \log(d_k)$. Now, choosing \(k=\lfloor d/2\rfloor+1\) gives $I(A:B|E)_{\rho^{(k)}} = O(d^{-1})$,
so the conditional mutual information tends to zero as \(d\to\infty\).

We now show that \(\rho_{ABE}^{(k)}\) is nevertheless bounded away from the set of exact quantum Markov chains.
Let \(\sigma_{ABE}\) be any exact Markov chain, so \(I(A:B|E)_\sigma=0\). Exact quantum Markov states have the structure theorem \cite{hayden_structure_2004}; in particular, the marginal \(\sigma_{AB}\) is separable across \(A:B\). 
For any pure product state \(|a\rangle\otimes |b\rangle\),
\begin{align}
    \langle ab|P_-|ab\rangle
=
\frac{1-|\langle a|b\rangle|^2}{2}
\le \frac12
\end{align}
By convexity, for every separable state \(\tau_{AB}\), $\operatorname{Tr}(P_-\tau_{AB})\le \frac12$.
On the other hand, $\operatorname{Tr}(P_-\rho^{(k)}_{AB})=1$.
Using the variational formula for trace distance, we have
\begin{align}
    \frac12\|\rho^{(k)}_{AB}-\tau_{AB}\|_1
\ge
\operatorname{Tr}(P_-\rho^{(k)}_{AB})-\operatorname{Tr}(P_-\tau_{AB})
\ge \frac12
\end{align}
Hence, for every exact Markov chain \(\sigma_{ABE}\),
\begin{align}
    \frac12\|\rho_{ABE}^{(k)}-\sigma_{ABE}\|_1
\ge
\frac12\| \rho^{(k)}_{AB}-\sigma_{AB}\|_1 \ge \frac12
\end{align}
where we used monotonicity of trace distance under partial trace.
Thus \(\rho_{ABE}^{(k)}\) remains a constant distance away from the set of exact Markov states, even though \(I(A:B|E)_{\rho^{(k)}}\to 0\).

\subsection{Canonical purifications and the reduced density matrix on \(AA'\)}

Let \(\{|J\rangle: J\subset\{1,\dots,d\}, |J|=m\}\) be an orthonormal basis of \(\bigwedge^m(\mathbb{C}^d)\), indexed by \(m\)-element subsets:
\begin{align}
    |J\rangle
=
|i_1\wedge i_2\wedge \cdots \wedge i_m\rangle
:=
\frac{1}{\sqrt{m!}}
\sum_{\pi\in S_m}\operatorname{sgn}(\pi)\,
|i_{\pi(1)}\rangle\otimes\cdots\otimes |i_{\pi(m)}\rangle
\in \bigwedge\nolimits^m(\mathbb{C}^d).
\end{align}
If \(x\in J\), then \(J\setminus \{x\}\) is an \((m-1)\)-element subset, and $|J\setminus\{x\}\rangle
\in \bigwedge\nolimits^{m-1}(\mathbb{C}^d)$
denotes the corresponding normalized antisymmetric basis vector. Define $\varepsilon(J,x = i_r):=(-1)^{r-1}$. Then, one has the decomposition
\begin{align}
    |J\rangle
=
\frac{1}{\sqrt m}\sum_{x\in J}\varepsilon(J,x)\,
|x\rangle_A\otimes |J\setminus\{x\}\rangle_R
\end{align}
where \(A\) denotes the first tensor factor and \(R\) denotes the remaining \(m-1\) tensor factors. Indeed, this is just the standard expansion of an antisymmetric tensor by separating out the first slot.
The canonical purification of the maximally mixed state on \(\bigwedge^m(\mathbb{C}^d)\) is
\begin{align}
    |\psi_m\rangle_{AR\,A'R'}
=
\frac{1}{\sqrt{d_m}}
\sum_{|J|=m}
|J\rangle_{AR}\otimes |J\rangle_{A'R'}
\end{align}
where we choose the basis to be real.
Define
\begin{align}
    \omega_{AA'}^{(m)}:=\operatorname{Tr}_{RR'}|\psi_m\rangle\langle\psi_m|.
\end{align}
By \(U(d)\)-covariance and Shur's lemma, $\omega_{AA'}^{(m)}$ must be an isotropic state:
\begin{align}
    \omega_{AA'}^{(m)}
=
f_m \Phi_{AA'}
+
\frac{1-f_m}{d^2-1}\bigl(I-\Phi_{AA'}\bigr)
\end{align}
where
\begin{align}
    \Phi_{AA'}:=|\Omega\rangle\!\langle\Omega|,
\qquad
|\Omega\rangle=\frac{1}{\sqrt d}\sum_{i=1}^d |i\rangle_A\otimes |i\rangle_{A'}
\end{align}
and
\begin{align}
    f_m=\langle \Omega|\omega_{AA'}^{(m)}|\Omega\rangle
\end{align}
Taking $m=k$ and $m=k-1$ gives the reduced density matrix on $AA'$ for canonical purifications of $\rho^{(k)}_{ABE}$ and $\rho^{(k)}_{AE}$, respectively.
Substituting the expansion of \(|J\rangle\) into \(|\Psi_m\rangle\), we obtain
\begin{align}
    |\psi_m\rangle
=
\frac{1}{\sqrt{d_m}}\frac{1}{m}
\sum_{|J|=m}\sum_{x,y\in J}
\epsilon(J,x)\epsilon(J,y)\,
|x\rangle_A\otimes |y\rangle_{A'}
\otimes
|J\setminus\{x\}\rangle_R
\otimes
|J\setminus\{y\}\rangle_{R'}
\end{align}
Contracting with \(\langle \Omega|_{AA'}\), we find
\begin{align}
    \langle \Omega|_{AA'}|\psi_m\rangle
&=
\frac{1}{\sqrt{d_m}}\frac{1}{m\sqrt d}
\sum_{|J|=m}\sum_{x\in J}
|J\setminus\{x\}\rangle_R
\otimes
|J\setminus\{x\}\rangle_{R'} \nonumber\\
& = \frac{d-m+1}{\sqrt{d_m}\,m\sqrt d}
\sum_{|J'|=m-1}
|J'\rangle_R\otimes |J'\rangle_{R'},
\end{align}
where $J':= J \setminus \{x\}$.
Taking the squared norm of both sides, we get
\begin{align}
    f_m
=
\left\|
(\langle \Omega|_{AA'}\otimes I_{RR'})|\Psi_m\rangle
\right\|^2
=
\frac{(d-m+1)^2}{d_m\,m^2\,d}\,d_{m-1} = \frac{d-m+1}{dm}.
\end{align}
Substituting this into the decomposition of $\omega_{AA'}^{(m)}$ yields
\begin{align}
    \frac12\|\omega_{AA'}^{(k)}-\omega_{AA'}^{(k-1)}\|_1
=
\left|
\frac{d-k+1}{dk}
-
\frac{d-k+2}{d(k-1)}
\right|
=
\frac{d+1}{d\,k(k-1)}
\end{align}
For the choice \(k=\lfloor d/2\rfloor+1\), this is \(O(d^{-2})\), so
\begin{align}
    \|\omega_{AA'}^{(k)}-\omega_{AA'}^{(k-1)}\|_1 \to 0
\qquad (d\to\infty)
\end{align}
Thus, the reduced density matrices on \(AA'\) of the canonical purifications of \(\rho_{ABE}^{(k)}\) and \(\rho_{AE}^{(k)}\) become asymptotically identical.

\section{Toric code under decoherence}\label{app:toric code}

\subsection{The model}

Consider the toric code ground state living on the edges of a square lattice $\Lambda$
\begin{align}
    \rho_{0} = \prod_{\square \in \Lambda} \frac{1+A_{\square}}{2} \prod_{+  \in \Lambda}\frac{1+B_{+}}{2},
\end{align}
where $\square$ ($+$) denotes the plaquette (vertice), and $A_\square = \prod_{i\in \square} Z_i$, $B_+ = \prod_{i\in +} X_i$. Since every term in the product is a projection and commutes with each other, $\rho_0$ is a pure state. In fact, the reduced density matrix, $\rho_{0,Q} :=\tr_{\Lambda\backslash Q}(\rho_0)$, in any region $Q \subseteq \Lambda$ is also pure.

By adding a dephasing channel $\mathcal{E}_p(\rho):= (1-p)\rho + p X \rho X$ with $0\leq p\leq 1/2$ to each qubit, we obtain a mixed state 
\begin{align}
    \rho_p &= \sum_{m} \mathrm{Pr}(m) \Pi_m,\nonumber\\
    \mathrm{Pr}(m) & = \sum_{e} p^{|e|} (1-p)^{|\Lambda| - |e|}\delta(m = \partial e),\nonumber\\
    \Pi_m & = \prod_{i\in e,\partial e = m} X_i \rho_0 X_i = \prod_{\square} \frac{1+(-1)^{m_\square}A_\square}{2} \prod_+ \frac{1+B_+}{2}, 
\end{align}
where $m\in\{0,1\}^\Lambda$ is a binary vector living on the plaquettes, and $e\in\{0,1\}^\Lambda$ is a binary vector living on the edges subject to the constraint that the activated edges must connect the activated  plaquettes, which is captured by the delta function $\delta(m = \partial e)$. The physical meaning is that the dephasing channel will activate/flip edges associated with $A_\square$ creating the so-called m-anyons that corresponds to the configuration generated by the vector $m$. These m-anyons states are captured by the projection $\frac{1-A_\square}{2}$, and that changes the toric code ground state into the excited state $\Pi_m$, which is again a projection. Thus, the canonical purification of the decohered state $\rho_p$ is given by
\begin{align}
    |\psi_{\rho_{p}} \rangle &= \sum_{m} \sqrt{\mathrm{Pr}(m)} | m\rangle | m\rangle
\end{align}
where $| m\rangle | m\rangle:= \Pi_{m}\otimes I|\Omega\rangle$ with $|\Omega\rangle\propto \sum_s |s\rangle |s\rangle$. Interestingly, for each anyon configuration, $\mathrm{Pr}(m)$ is proportional to the partition function of the random bond Ising model (RBIM) \cite{noauthor_topological_nodate}:
\begin{align}
    \mathrm{Pr}(m) = c_0 Z[\eta]\delta(m = \partial \eta),\quad Z[\eta]:= \sum_{\{\sigma\}} e^{J \sum_{\langle \mu,\nu\rangle}\eta_{\mu\nu}\sigma_\mu \sigma_\nu} 
\end{align}
where $c_0$ is a constant that does not depend on $m$, $J = \frac{1}{2} \log(p/(1-p))$, $\sigma_\mu$ are Ising spins living on the plaquettes, and $\eta_{\mu \nu} := 2 e_i -1$ with $\mu\nu$ the two plaquettes that share the edge $e_i$. Notice that even though the choice of $\eta$ is not unique as $\eta \to \eta+c$ with $\partial c = 0$ results in the same $m$, different choices are already taken into account by combining spin configurations.

First, consider the two-point $XX$ correlation function. Due to orthogonality and translation invariance, 
\begin{align}
    C_{XX}(r = |i-j|):=&\langle\psi_{\rho_p}| (X_i \otimes X_i)(X_j \otimes X_j) |\psi_{\rho_p} \rangle - \langle\psi_{\rho_p}| X_i \otimes X_i |\psi_{\rho_p} \rangle \langle\psi_{\rho_p}| X_j \otimes X_j |\psi_{\rho_p} \rangle \nonumber\\
    =& \sum_m \sqrt{\mathrm{Pr}(m) \mathrm{Pr}(m'' = \partial(e+e_i+e_j)) } - \left(\sum_m \sqrt{\mathrm{Pr}(m) \mathrm{Pr}(m' = \partial(e+e_i)) }\right)^2,
\end{align} 
Using the following identity
\begin{align}
    \sum_m \sqrt{\mathrm{Pr}(m) \mathrm{Pr}(m') } = \sum_m \mathrm{Pr}(m) \sqrt{\frac{\mathrm{Pr}(m')}{\mathrm{Pr}(m)}} = \mathbb{E}_{m\sim \mathrm{Pr}(m)} \sqrt{\frac{Z[\eta']}{Z[\eta]}}|_{m=\partial\eta},
\end{align}
we can rewrite the correlation function as
\begin{align}\label{eq:CXX}
    C_{XX}(r) &= \mathbb{E}_{m\sim \mathrm{Pr}(m)}\sqrt{\frac{Z[\eta+e_i+e_j]}{Z[\eta]}}|_{m=\partial\eta} - \left(\mathbb{E}_{m\sim \mathrm{Pr}(m)}\sqrt{\frac{Z[\eta+e_i]}{Z[\eta]}}|_{m=\partial\eta}\right)^2 
    % & = \mathbb{E}_{m\sim \mathrm{Pr}(m)}\sqrt{\frac{Z[\eta+e_i]}{Z[\eta]}} \sqrt{\frac{Z[\eta+e_j]}{Z[\eta]}}|_{m=\partial\eta} - \left(\mathbb{E}_{m\sim \mathrm{Pr}(m)}\sqrt{\frac{Z[\eta+e_i]}{Z[\eta]}}|_{m=\partial\eta}\right)^2
\end{align}
where $\eta+e_i$ is understood to flip the sign of $\eta_{\mu\nu}$ perpendicular to the edge $e_i$.
% and in the second equation we use $\sum_m \sqrt{\mathrm{Pr}(m) \mathrm{Pr}(m+m_1 + m_2)} =\sum_m \sqrt{\mathrm{Pr}(m+m_1) \mathrm{Pr}(m+ m_2)}$. Moreover, the ratio between partition functions can be regarded as the expectation value of a local bond-flipping operator, which leads to
% \begin{align}
%     C_{XX}(r)  = \mathbb{E}_{m\sim \mathrm{Pr}(m)}\sqrt{\langle \mathcal{O}_i\rangle_\eta \langle \mathcal{O}_j\rangle_\eta} |_{m=\partial\eta} - \left(\mathbb{E}_{m\sim \mathrm{Pr}(m)}\sqrt{\langle \mathcal{O}_i\rangle_\eta}|_{m=\partial\eta}\right)^2,\quad  \mathcal{O}_i:= e^{-2J \eta_{\mu\nu}\sigma_\mu \sigma_\nu},
% \end{align}
% where $\eta_{\mu\nu}$ is perpendicular to $e_i$. In this way, we find that the correlation function $C_{XX}$ is similar to the Edward-Anderson order parameter in spin glass theory \cite{}. Since the RBIM has no thermodynamic glass order at finite temperature \cite{bhatt_numerical_1988,hartmann_lower_2001, katzgraber_correlation_2004}.

\begin{figure}[tp]
  \centering
  \includegraphics[width=0.8\textwidth]{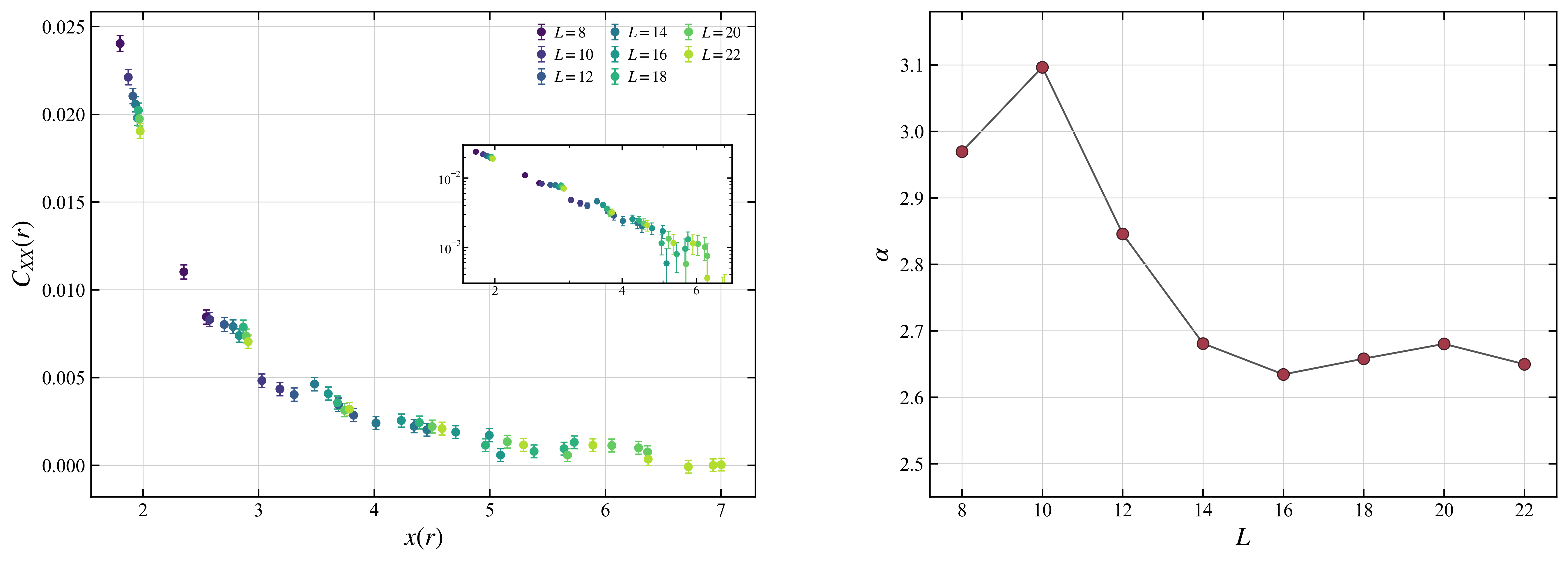}
  \caption{Decay of the connected two-point correlation function near critical point $p_c = 0.109$. Left: $C_{XX}$ decays in power-law (inset) in the chord distance $x(r):= \frac{L}{\pi}\sin\frac{\pi r}{L}$ for eight different system sizes $L$. Right: the exponent $\alpha$ from fitting $C_{XX}(r)\sim x(r)^{-\alpha}$ approaches a constant when increasing the system size. The data are averaged over $10^6$ samples.}
  \label{fig:connect}
\end{figure}

To analyze the critical behavior of $C_{XX}(r)$, we expand the first line of \eqref{eq:CXX} as follows (for simplicity, we do not include subscripts from now on)
\begin{align}
    C_{XX}(r) &= \mathbb{E}\sqrt{\langle \mathcal{O}_i \mathcal{O}_j\rangle} -\left(\mathbb{E}\sqrt{\langle \mathcal{O}_i \rangle} \right)^2  \nonumber\\
    & = \mathbb{E}\sqrt{\langle \left(\cosh(2J) - \sinh(2J)A_i \right) \left(\cosh(2J) - \sinh(2J)A_j \right)\rangle} -\left(\mathbb{E}\sqrt{\langle \cosh(2J) - \sinh(2J)A_i  \rangle} \right)^2 \nonumber\\
    & = \cosh(2J)\left[ \mathbb{E}\sqrt{ \left(1 - \tanh(2J)\langle A_i \rangle  \right)^2  + \tanh(2J)^2 \langle A_iA_j \rangle_{\mathrm{c}} } -\left(\mathbb{E}\sqrt{1 - \tanh(2J)\langle A_i  \rangle} \right)^2 \right]
\end{align}
where $A_i := \eta_{\mu\nu}\sigma_\mu \sigma_\nu$ with $i$ denoting the edge $e_i$ perpendicular to $\eta_{\mu\nu}$. Near the critical point, the connected correlation $\langle A_i A_j\rangle_{\mathrm{c}}:= \langle A_i A_j\rangle - \langle A_i\rangle \langle A_i\rangle$ will be small, so the taylor expansion gives
\begin{align}
    C_{XX}(r)\approx \cosh(2J) \left[\Delta A_i + \mathbb{E} \frac{\tanh(2J)^2 \langle A_i A_j\rangle_{\mathrm{c}}}{2 \left(1 - \tanh(2J)\langle A_i \rangle  \right)}\right],
\end{align}
where $\Delta A_i:=  1 - \tanh(2J)\mathbb{E}\langle A_i  \rangle -\left(\mathbb{E}\sqrt{1 - \tanh(2J)\langle A_i  \rangle} \right)^2$. Therefore, we expect that the power-law decay of $C_{XX}$ near the critical point will come from the power-law decay of $\langle A_i A_j\rangle_{\mathrm{c}}$. Shown in Fig.~\ref{fig:connect} is the critical behavior of $C_{XX}(r)$ near $p_c = 0.109$. The inset plot on the left indicates that $C_{XX}(r)$ decays in power-law in distance with nearly zero offset $\Delta A_i\approx 0$\footnote{Since $\Delta A_i$ can be considered as the Edward-Anderson order parameter, $\Delta A_i=0$ is consistent with the absence of the spin glass order at finite temperature \cite{bhatt_numerical_1988,hartmann_lower_2001, katzgraber_correlation_2004}.}. By fitting the power-law exponent $C_{XX}(r) \sim x(r)^{-\alpha}$, we find $\alpha \approx 8/3$, which is consistent with the scaling dimension for the energy density operators reported in \cite{Delfino2025Nishimori}. However, as $C_{XX}$ rapidly decays to zero, the errors at a large distance are significant, so we would like to leave conclusive results to future work.

Next, we look at the expectation value of the string operators. There are two kinds of string operators; one involves an open string with two ends, and the other one is the logical operator which involves a non-contractible loop. They are defined as
\begin{align}
    C_{\mathrm{dis}}(|\gamma|=L/2):=& \langle\psi_{\rho_p}| \prod_{i\in \gamma}(X_i \otimes X_i) |\psi_{\rho_p} \rangle \\
    C_{\mathrm{logical}}( \gamma: \text{non-contractible loop} ):=& \langle\psi_{\rho_p}| \prod_{i\in \gamma,\partial\gamma=0}(X_i \otimes X_i) |\psi_{\rho_p} \rangle, 
\end{align}
where $\gamma$ is a path in the dual lattice and $i\in \gamma$ means the edges $e_i$ that $\gamma$ transverse. Using the above mapping, we have 
\begin{align}
    C_{\mathrm{dis}} =&  \mathbb{E}_{m\sim \mathrm{Pr}(m)}\sqrt{\frac{Z[\eta+\sum_{i\in\gamma} e_i]}{Z[\eta]}}|_{m=\partial\eta} \\
    C_{\mathrm{logical}} =&  \mathbb{E}_{m\sim \mathrm{Pr}(m)}\sqrt{\frac{Z[\eta+\sum_{i\in\gamma,\partial\gamma=0} e_i]}{Z[\eta]}}|_{m=\partial\eta}
\end{align}

In numerics, we take $M$ samples by randomly flipping edges with probability $p$, and, for each sample, we compute the partition function $Z[\eta]$ using the matchgate tensor network, which we describe in the following.

\subsection{The matchgate tensor network algorithm}
The purpose of this subsection is to compute the RBIM partition function on a graph $G=(V,E)$:
\begin{align}
    Z(\tau ; \beta)=\sum_{\sigma \in\{ \pm 1\}^V} \exp \left(\beta \sum_{e=\langle i j\rangle \in E} \tau_e \sigma_i \sigma_j\right)
\end{align}
Let $G$ be an $L\times L$ square lattice with periodic boundary conditions. The idea is that the computation of such a partition function can be reduced to a problem called matching that is equivalent to computing a Pfaffian according to the Fisher–Kasteleyn–Temperley algorithm \cite{Kasteleyn1961,TemperleyFisher1961,fisher}, and the matchgate tensor network is a family of tensor networks that can solve these Pfaffians \cite{Cai,Bravyi2009MatchgateNonplanar}
. In the following, we will follow \cite{Bravyi2009MatchgateNonplanar} to perform the numerics.

The partition function can be rewritten as a weighted summation by expanding the exponential function to $\exp(K_e \sigma_i \sigma_j) = \cosh K_e(1+\tanh K_e \sigma_i \sigma_j)$. Let $K_e:=\beta \tau_e$. The partition function is nonzero only if every vertex has even incident edges; otherwise, adding up $\sigma_i$ gives zero. This leads to
\begin{align} 
&Z=2^{|V|} \prod_{e \in E} \cosh K_e \cdot Z_{\text {even }}(x), \quad Z_{\text {even }}(x)=\sum_{\substack{\gamma \subseteq E \\ \partial \gamma=0}} \prod_{e \in \gamma} x_e,\quad x_e:=\tanh K_e = \tau_e \tanh \beta 
\end{align}
We call $Z_{\text{even}}$ the partition function in an even subgraph, and we call such an expansion a high-$T$ expansion. Introduce an occupation bit $n_{v,e}$ on each half-edge incident to vortex $v$. The local vertex tensor is
\begin{equation}
 T_v(n_0,n_1,n_2,n_3)
 =\begin{cases}
 1,&n_0+n_1+n_2+n_3=0\pmod 2,\\
 0,&n_0+n_1+n_2+n_3=1\pmod 2,
 \end{cases}
 \label{eq:even-tensor}
\end{equation}
An ordinary weighted tensor-network expression is
\begin{equation}
 Z_{\mathrm{even}}(x)
 =\sum_{\{n_{v,e}\}}
 \prod_{v\in V}T_v(\{n_{v,e}\}_{e\ni v})
 \prod_{e=(u,v)}\delta_{n_{u,e},n_{v,e}}x_e^{n_{u,e}}.
 \label{eq:ordinary-TN}
\end{equation}
According to \cite{Bravyi2009MatchgateNonplanar}, this partition function on a torus can be represented by four $4L^2\times 4L^2$ skew matrix $K^{(a,b)}$ labeled by the four boundary conditions $(a,b)\in\{0,1\}^2$. The matrix $K^{(a,b)}$ is built as follows. Let $I(r,c,\ell)=4(rL+c)+\ell$ be the global index: $(r,c)$ is the vortex index, $\ell$ labels the $4\times 4$ skew matrix $A_0$, which is given by
\begin{equation}
 A_0=
 \begin{pmatrix}
 0&1&1&1\\
 -1&0&1&1\\
 -1&-1&0&1\\
 -1&-1&-1&0
 \end{pmatrix}.
 \label{eq:Astar}
\end{equation}
Then, we assign
\begin{align}
    \text{local blocks:}\quad  K^{(a,b)}_{I(r,c,i),I(r,c,j)} & = (A_0)_{ij} \\
    \text{horizontal bonds:}\quad  K^{(a,b)}_{I(r,c,1),I(r,c+1,3)} & = (-1)^{a \mathbf{1}_{c=L-1}} x^h_{r,c} \\
    \text{vertical bonds:}\quad  K^{(a,b)}_{I(r,c,2),I(r+1,c,0)} & = -(-1)^{b \mathbf{1}_{r=L-1}} x^v_{r,c}
\end{align}
where $x^h_{r,c}:= x_e \delta_{e,((r,c),(r,c+1))}$ and $x^v_{r,c}:= x_e \delta_{e,((r,c),(r+1,c))}$. After assignment, $K^{(a,b)}$ is antisymmetrized to become a skew matrix. The Pfaffian of $K^{(a,b)}$ is then related to the partition function in each winding sector $(h_x,h_y) \in\{0,1\}^2$ through Kasteleyn signs \cite{ChelkakCimasoniKassel2017IsingCombinatorics}
\begin{align}
P_{ab}:=\mathrm{Pf}(K^{(a,b)})=\sum_{h_x,h_y}
 (-1)^{q_0(h_x,h_y)+a h_x+b h_y}Z_{h_xh_y},\quad q_0(h_x,h_y)=h_x+h_y+h_xh_y\pmod2 
 \end{align}
 Therefore, the even-graph partition function is given by
 \begin{align}
     Z_{\mathrm{even}}= Z_{00}+Z_{10}+Z_{01}+Z_{11} = \frac{-P_{00}+P_{01}+P_{10}+P_{11}}{2},
 \end{align}
 which involves four Pfaffian computations.
 An efficient way to compute the Pfaffian is the Schur contraction. The idea is to repeatly use the identity
 \begin{align}
     \mathrm{Pf}\begin{pmatrix}A&B\\-B^{ T}&D\end{pmatrix}
 =\mathrm{Pf}(A)\mathrm{Pf}\!\left(D+B^{ T}A^{-1}B\right)
 \end{align}
 for an invertible skew matrix $A$ to reduce the dimension of the entire matrix. A standard manipulation gives a time cost of order $O(L\times L^3)$, where $O(L^3)$ is for a $L\times L$ skew matrix; by contrast, a direct Pfaffian computation costs $\mathrm{time}\sim O(L^6)$.

 There is another way of expanding the partition function into a matchgate tensor network. Define the domain wall operator and the weight
 \begin{align}
     \eta_e=\frac{1-\sigma_i\sigma_j}{2}\in\{0,1\},\quad q:=e^{-2\beta},\quad w_e:= q^{\tau_e}
 \end{align}
 Then, the partition function can be written as a weighted summation on the dual graph $G^\star$:
 \begin{align}
     Z = \prod_e e^{\beta \tau_e} \sum_{\{\eta\}} \prod_e w_e^{  \eta_e} =2e^{\beta|E|}q^{n_-}
 \sum_{\substack{D\text{ even on }G^\star\\ {}[D]=0}}
 \prod_{e\in D}w_e,
 \end{align}
 where $n_-$ is the number of edges with $\tau_e = -1$. The summation is over even subgraph in $G^\star$ because the domain wall cannot end, and it must be in the trivial twist sector because the spin configuration is globally defined. The overall factor of $2$ comes from the global spin flip. We call this the low-$T$ expansion. We can contrast the low-$T$ and high-$T$ expansion by noting that when $\beta\to \infty$, the weight of high-$T$ expansion is large $x_e \to 1$, while the weight of low-$T$ expansion is small $q\to 0$. Therefore, it is more favorable to use low-$T$ expansion at low $T$ to avoid large numbers whose signed combination is much smaller. A priori, however, both high-$T$ and low-$T$ expansions are exact.
 Now, the low-$T$ Pfaffian network $\tilde K^{(a,b)}$ lives in the dual lattice. Local blocks are still given by the skew matrix $A_0$, while horizontal/vertical bonds are given by replacing $x^{h/v}_{r,c}\rightarrow q^{\tau^{v/h}_{r,c}}$, where $\tau^{v/h}$ is defined similarly as $x^{h/v}$; note that a horizontal bond in $G$ corresponds to a vertical bond in $G^\star$. Then, the total partition function is given by a single winding sector:
 \begin{align}
     Z=\tilde Z_{00}
 =2e^{\beta|E|}q^{n_-} \frac{\tilde P_{00}+ \tilde P_{10}+\tilde P_{01}+\tilde P_{11} }{4},
 \end{align}
 with $\tilde P_{ab}:= \mathrm{Pf}(\tilde K^{(a,b)})$.

\section{Free-fermion simulation of the dephased Chern insulator}
\label{app:chern-numerics}

In this appendix we describe the numerical methods used for the Chern-insulator
example in the main text.  The calculation has two complementary parts.  For
partial dephasing, $0\leq p<1$, we unravel the flagged channel into number
measurements.  Every trajectory is then a Slater determinant, which makes both
the conditional mutual information (CMI) and observables in the canonical
purification accessible by Monte Carlo sampling.  At the fully measured
endpoint, $p=1$, the state becomes fully classical, which allows for a more direct estimator and lets us reach systems as large as
$48\times48$ sites.

\subsection{Model and noise convention}
\label{app:chern-model}

We use the Hofstadter model in Landau gauge,
\begin{equation}
 H=\sum_{x,y}\left(
 c_{x+1,y}^{\dagger}c_{x,y}
 +e^{2\pi i x/q}c_{x,y+1}^{\dagger}c_{x,y}
 +\mathrm{h.c.}\right),
 \label{eq:app-hofstadter}
\end{equation}
at filling $1/q$.  With the hopping and flux convention in
Eq.~\eqref{eq:app-hofstadter}, the lowest band has Chern number $C=-1$.
We used $q=4$ for the CMI calculations and $q=5$ for the partial-translation
diagnostic.  These choices describe the same $|C|=1$ phase; they were selected
for numerical convenience in their respective geometries.

The noise channel is the flagged dephasing in number basis. Let $p$ be the probability that a site is projectively measured, the corresponding unflagged channel is
\begin{align}
 \mathcal N^{[i]}_p(\rho)
 &= (1-p)\rho+p\left[n_i\rho n_i+(1-n_i)\rho(1-n_i)\right]\nonumber\\
 &=\left(1-\frac p2\right)\rho
   +\frac p2(1-2n_i)\rho(1-2n_i).
 \label{eq:app-dephasing-convention}
\end{align}
In this convention, the fully measured
endpoint discussed below is $p=1$.

We consider the measurement channel that retains the flag of measurement outcomes,
\begin{equation}
 \mathcal M^{[i]}_p(\rho)
 =\sum_{m_i=0}^{2}w_{m_i}P_{m_i}\rho P_{m_i}
   \otimes\ket{m_i}\!\bra{m_i}_{F_i},
 \label{eq:app-fine-channel}
\end{equation}
where
\begin{equation}
 P_0=\mathbf 1,\qquad P_1=n_i,\qquad P_2=1-n_i,
 \qquad w_0=1-p,\quad w_1=w_2=p.
\end{equation}
The three flag values mean ``unmeasured,'' ``measured occupied,'' and
``measured empty,'' respectively.  Coarse-graining the last two outcomes gives
the flagged dephasing channel used in the main text; and the two flagged
descriptions are related by a locally reversible channel.  After applying
Eq.~\eqref{eq:app-fine-channel} to all sites, the state takes the block-diagonal
form
\begin{equation}
 \rho_p=\sum_{\boldsymbol m}P(\boldsymbol m)
 \rho_{\boldsymbol m}\otimes
 \ket{\boldsymbol m}\!\bra{\boldsymbol m}_{F},
 \label{eq:app-trajectory-ensemble}
\end{equation}
with
\begin{align}
 P(\boldsymbol m)
 &=p^{|\boldsymbol m|}(1-p)^{N-|\boldsymbol m|}
   \bra{\psi_0}P_{\boldsymbol m}\ket{\psi_0},\nonumber\\
 \rho_{\boldsymbol m}
 &=\frac{P_{\boldsymbol m}\ket{\psi_0}\!\bra{\psi_0}
 P_{\boldsymbol m}}{\bra{\psi_0}P_{\boldsymbol m}\ket{\psi_0}}.
 \label{eq:app-trajectory-probability}
\end{align}
Here $|\boldsymbol m|$ counts measured sites, and
$P_{\boldsymbol m}$ is the product of the corresponding number projectors.
The Born probability in Eq.~\eqref{eq:app-trajectory-probability} is generated
sequentially, so no exponentially large probability table is constructed.

\subsection{Gaussian trajectory method}
\label{app:chern-gaussian}
\subsubsection{Correlation and entanglement}
A number-conserving Slater determinant with $N_f$ occupied orbitals is stored
as an $N\times N_f$ isometry $Q$ with $Q^{\dagger}Q=\mathbf 1$.  Its
correlation matrix is $G=QQ^{\dagger}$.  This representation gives a compact formula
for the state overlap.  If $u$ is a one-particle unitary and $\widehat u$ its
second-quantized operator, then
\begin{equation}
 \bra{Q_2}\widehat u\ket{Q_1}
 =\det\!\left(Q_2^{\dagger}uQ_1\right).
 \label{eq:app-slater-overlap}
\end{equation} 
Furthermore, if $R$ is a set of sites, the entropy
of the reduced state is
\begin{equation}
 S_R=-\sum_a\left[\nu_a\log\nu_a+(1-\nu_a)\log(1-\nu_a)\right],
 \label{eq:app-gaussian-entropy}
\end{equation}
where $\{\nu_a\}$ are the eigenvalues of $G_R$, the restriction of $G$ to
$R$. 

\subsubsection{Efficient implementation of number basis measurement}
Projecting the occupation number of site $i$ preserves Gaussianity.  The two Born
probabilities are
\begin{equation}
 P(n_i=1)=G_{ii},\qquad P(n_i=0)=1-G_{ii}.
\end{equation}
We now describe the corresponding orbital update.  Let $e_i$ be the
one-particle state localized at site $i$, and let
\begin{equation}
 r_i=e_i^\dagger Q,
 \qquad g_i=r_i r_i^\dagger=G_{ii}.
 \label{eq:app-measurement-row}
\end{equation}
Thus $r_i$ is the $i$th row of $Q$ and specifies which linear combination of
the occupied orbitals has support on the measured site.

For the occupied outcome, $n_i=1$, one orbital becomes pinned to $e_i$.  The
remaining $N_f-1$ occupied orbitals vanish at site $i$.  To construct
them, choose an $N_f\times(N_f-1)$ matrix $B_i$ satisfying
\begin{equation}
 r_iB_i=0,
 \qquad B_i^\dagger B_i=\mathbf 1_{N_f-1}.
 \label{eq:app-occupied-nullspace}
\end{equation}
An orthonormal orbital frame for the normalized post-measurement state is then
\begin{equation}
 Q^{(1)}_i=\left(e_i,\,QB_i\right).
 \label{eq:app-occupied-update}
\end{equation}
Indeed, $QB_i$ is orthonormal because $Q^\dagger Q=\mathbf 1$, and it is
orthogonal to $e_i$ because $r_iB_i=0$. decomposition. We obtain the isometry $B_i$ in the following way. Let $x_i=\frac{r_i^\dagger}{\sqrt{g_i}}$
be the normalized column vector associated with the $i$th row of $Q$. We
construct a Householder reflector $H_i$ such that
\begin{equation}
    H_i x_i=e^{i\phi_i}f_1,
    \label{eq:app-householder-action}
\end{equation}
for some unimportant phase $\phi_i$, where $f_1 = (1, 0, 0,...,0)^{T}\in \mathbb{C}^{N_f}$. Writing
$H_i=(h_1,h_2,\ldots,h_{N_f})$ in terms of its columns, the matrix $B_i$ is constructed as
\begin{equation}
    B_i=(h_2,\ldots,h_{N_f}).
\end{equation}
Because a Householder reflector is Hermitian and unitary,
Eq.~\eqref{eq:app-householder-action} implies
\begin{equation}
    r_i B_i
    =\sqrt{g_i}\,x_i^\dagger B_i=0,
    \qquad
    B_i^\dagger B_i=\mathbf 1_{N_f-1}.
\end{equation}
Thus, the columns of $B_i$ form an orthonormal basis for the null space of
$r_i$, as required by Eq.~\eqref{eq:app-occupied-nullspace}. Note that we have avoided directly computing the nullspace of $r_i$ because that costs $O(N^3_f)$.

For the empty outcome, $n_i=0$, first project every occupied orbital away from
site $i$,
\begin{equation}
 \widetilde Q^{(0)}_i=\Pi_i Q,
 \qquad \Pi_i=\mathbf 1-e_ie_i^\dagger.
\end{equation}
These columns span the post-measurement occupied subspace but are no
longer orthonormal. We then re-orthonormalize them semi-analytically. Their Gram matrix differs from the identity by one rank,
\begin{equation}
 \widetilde Q_i^{(0)\dagger}\widetilde Q_i^{(0)}
 =\mathbf 1-r_i^\dagger r_i.
\end{equation}
Consequently, the normalized orbital frame can be written explicitly as
\begin{align}
 Q_i^{(0)}
 &=\Pi_iQ\left(\mathbf 1-r_i^\dagger r_i\right)^{-1/2}\nonumber\\
 &=\Pi_iQ\left[
 \mathbf 1+\frac{(1-g_i)^{-1/2}-1}{g_i}
 r_i^\dagger r_i\right].
 \label{eq:app-empty-update}
\end{align}
The second line follows because $r_i^\dagger r_i$ has only one nonzero
eigenvalue, $g_i$.
Equations~\eqref{eq:app-occupied-update} and
\eqref{eq:app-empty-update} therefore replace a general QR reorthogonalization by more efficient
Householder and rank-one updates. 

Applying these updates in a fixed site order gives the conditional Born
probability at every step.  Drawing each outcome with that probability and
multiplying the conditional probabilities samples
Eq.~\eqref{eq:app-trajectory-probability} exactly.

\subsubsection{Canonical purification}
The trajectory representation also gives a direct rule for
canonical-purification observables.  If operators $O_1$ and $O_2$ are diagonal in the flags,
then
\begin{align}
 \bra{\sqrt{\rho_p}}O_1\otimes O_2
 \ket{\sqrt{\rho_p}}\nonumber=\sum_{\boldsymbol m}P(\boldsymbol m)
 \bra{\psi_{\boldsymbol m}}O_1\ket{\psi_{\boldsymbol m}}
 \bra{\psi^*_{\boldsymbol m}}O_2\ket{\psi^*_{\boldsymbol m}}.
 \label{eq:app-cp-diagonal-observable}
\end{align}
The two Gaussian matrix elements are evaluated on the same trajectory.   

More generally, let $V_f$ be a permutation of the flag basis,
\begin{equation}
    V_f\ket{\boldsymbol m}_F
    =\ket{f(\boldsymbol m)}_F.
\end{equation}
For an observable
$\widehat{\mathcal O}
=O_1\otimes O_2\otimes V_f\otimes V_f$
acting on the two fermion copies and the two flag registers, respectively,
orthogonality of the flag states gives
\begin{align}
    \langle\widehat{\mathcal O}\rangle_{\sqrt{\rho}}
    ={}
    \sum_{\boldsymbol m}
    \sqrt{P(\boldsymbol m)P(f(\boldsymbol m))}
    \bra{\psi_{f(\boldsymbol m)}}O_1\ket{\psi_{\boldsymbol m}}
    \bra{\psi^*_{f(\boldsymbol m)}}O_2
    \ket{\psi^*_{\boldsymbol m}}.
\end{align}
Here we used
$\bra{\boldsymbol n}V_f\ket{\boldsymbol m}
=\delta_{\boldsymbol n,f(\boldsymbol m)}$.
Equivalently, sampling $\boldsymbol m$ from $P(\boldsymbol m)$ yields
\begin{align}
    \langle\widehat{\mathcal O}\rangle_{\sqrt{\rho}}
    =\mathbb E_{\boldsymbol m\sim P}\Bigg[
    \sqrt{\frac{P(f(\boldsymbol m))}
                 {P(\boldsymbol m)}}
    \bra{\psi_{f(\boldsymbol m)}}O_1\ket{\psi_{\boldsymbol m}}
    \bra{\psi^*_{f(\boldsymbol m)}}O_2
    \ket{\psi^*_{\boldsymbol m}}
    \Bigg].
\end{align}

\subsection{Conditional mutual information}
\label{app:chern-cmi}

For a region $R$, the orthogonal flag blocks imply
\begin{equation}
 S(RF_R)=H[P_R(\boldsymbol m_R)]
 +\sum_{\boldsymbol m_R}P_R(\boldsymbol m_R)
 S\!\left(\rho_{R|\boldsymbol m_R}\right).
 \label{eq:app-block-entropy}
\end{equation}
The first term is a Shannon entropy of the local record, while the second is
the average Gaussian entropy of the conditional fermion state. We can then compute the CMI through sampling measurement trajectories. The contribution of one sample is
\begin{align}
 \widehat I_{\boldsymbol m}(A:C|B)
 ={}&\log P_{ABC}(\boldsymbol m_{ABC})
   +\log P_B(\boldsymbol m_B)\nonumber\\
 &-\log P_{AB}(\boldsymbol m_{AB})
   -\log P_{BC}(\boldsymbol m_{BC})\nonumber\\
 &+S_{AB|\boldsymbol m_{AB}}+S_{BC|\boldsymbol m_{BC}}\nonumber\\
 &-S_{B|\boldsymbol m_B}-S_{ABC|\boldsymbol m_{ABC}}.
 \label{eq:app-coupled-cmi-estimator}
\end{align}
Its average is exactly $I(A:C|B)$.  Sharing the measurement record among all four terms is
important: their fluctuations are strongly correlated, and the coupled
estimator has much lower variance than subtracting four separately sampled
entropies.

\subsection{Partial translation and the Chern number of the canonical purification}
\label{app:chern-partial-translation}

We diagnose the topology of the canonical purification using the partial
translation construction of
Ref.~\cite{Kobayashi_2025}. That work combines a translation of
one half of a cylinder with a charge rotation and thereby separates the
higher chiral-central-charge and higher-Hall responses.  We first review its
specialization to an invertible Chern insulator and then adapt it to the doubled, flagged state.

Place the system on a cylinder that is open along $x$ and antiperiodic along
$y$, and let $A$ be its left half.  The antiperiodic boundary condition removes
the accidental zero-mode degeneracy.  Define $T_{A,2\pi/n}$ to translate the
sites in $A$ by $L_y/n$ lattice spacings around the cylinder, including the
minus sign whenever a fermion crosses the antiperiodic seam.  Also define
$U_{A,\phi}$ by
\begin{equation}
 U_{A,\phi}c_{x,y}U_{A,\phi}^{\dagger}=e^{-i\phi}c_{x,y},
 \qquad (x,y)\in A.
\end{equation}
Ref. ~\cite{Kobayashi_2025} introduces the expectation value of the partial-translation operator:
\begin{equation}
 \mathcal T(2\pi/n,s\pi/n)
 =\bra{\psi}U_{A,s\pi/n}T_{A,2\pi/n}\ket{\psi}.
\end{equation}
For odd $n$, the universal contribution is
\begin{equation}
    \mathcal T\left(\frac{2\pi}{n},\frac{s\pi}{n}\right)
    \propto
    \exp\left[
        \frac{2\pi i}{n}
        \left(
            \frac{s^2}{8}\sigma_H
            -\frac{n^2+2}{24}c_-
        \right)
    \right],
    \label{eq:app-partial-translation-universal}
\end{equation}
where the proportionality factor is real and positive. 
 
For an invertible integer
quantum Hall state, $\sigma_H=c_-=C$, and the result reduces to
\begin{equation}
 \arg\mathcal T(2\pi/n,s\pi/n)
 =\frac{2\pi}{n}\left[\frac{s^2}{8}
 -\frac{n^2+2}{24}\right]C
 \quad (\mathrm{mod}\ 2\pi),
 \label{eq:app-higher-hall-phase}
\end{equation}
up to corrections that vanish as the cylinder circumference is increased.
For $n=3$, the choices $s=1$ and $s=3$ give
\begin{equation}
 \arg\mathcal T(2\pi/3,\pi/3)=-\frac{2\pi}{9}C,
 \qquad
 \arg\mathcal T(2\pi/3,\pi)=\frac{4\pi}{9}C.
 \label{eq:app-single-copy-phases}
\end{equation}

The canonical purification of Eq.~\eqref{eq:app-trajectory-ensemble} is
\begin{equation}
 \ket{\sqrt{\rho_p}}=
 \sum_{\boldsymbol m}\sqrt{P(\boldsymbol m)}
 \ket{\psi_{\boldsymbol m}}\ket{\psi_{\boldsymbol m}^{*}}
 \ket{\boldsymbol m}_{F}\ket{\boldsymbol m}_{F'}.
 \label{eq:app-canonical-purification-trajectories}
\end{equation}
At $p=0$, the first copy has Chern number $C$, while the complex-conjugate copy
has Chern number $-C$.  Consequently, the total charge Hall response cancels,
but a spin-Hall response can be isolated by using different charge rotations on
the two copies.  Combining the two phases in
Eq.~\eqref{eq:app-single-copy-phases}, we obtain
\begin{equation}
 \arg\bra{\sqrt{\rho_p}}
 U_{A,\pi/3}U_{A',\pi}T_{AA',2\pi/3}
 \ket{\sqrt{\rho_p}}
 =-\frac{2\pi}{3}C.
 \label{eq:app-spin-hall-phase}
\end{equation}
The translation acts on both the fermions and their flags, whereas the charge
rotations act only on the fermions. 

Let $T\boldsymbol m$ denote the record obtained by translating the flags in
$A$.  Orthogonality of the flag states reduces the expectation value in
Eq.~\eqref{eq:app-spin-hall-phase} to
\begin{align}
 \mathcal T_{\mathrm{CP}}=
 \sum_{\boldsymbol m}\sqrt{P(\boldsymbol m)P(T\boldsymbol m)}
 \bra{\psi_{T\boldsymbol m}}U_{A,\pi/3}T_A
 \ket{\psi_{\boldsymbol m}} \bra{\psi^*_{T\boldsymbol m}}U_{A',\pi}T_{A'}
 \ket{\psi^*_{\boldsymbol m}}.
 \label{eq:app-cp-translation-sum}
\end{align}
Sampling $\boldsymbol m$ from $P(\boldsymbol m)$ turns this into the importance
sampling estimator
\begin{equation}
 \widehat{\mathcal T}_{\boldsymbol m}=
 \sqrt{\frac{P(T\boldsymbol m)}{P(\boldsymbol m)}}
 \bra{\psi_{T\boldsymbol m}}U_{A,\pi/3}T_A
 \ket{\psi_{\boldsymbol m}}
 \bra{\psi^*_{T\boldsymbol m}}U_{A',\pi}T_{A'}
 \ket{\psi^*_{\boldsymbol m}}.
 \label{eq:app-cp-translation-estimator}
\end{equation}
Both overlaps are determinants of the form
Eq.~\eqref{eq:app-slater-overlap}.  

We used $L_x=20$, $L_y=15,30,45$, flux $2\pi/5$, filling $1/5$, and a cut at
$x=10$.  Each point initially used $2500$ trajectories; the $L_y=30,45$ data
at $p=0.7,0.8,0.9$ were increased to $10^4$ trajectories.  We extracted
\begin{equation}
 C_{\mathrm{CP}}=-\frac{3}{2\pi}\arg\mathcal T_{\mathrm{CP}}.
\end{equation}
As shown in the main text, $C_{\mathrm{CP}}$ approaches $-1$ as $L_y$ grows
for every sampled $p<1$.  The magnitude of $\mathcal T_{\mathrm{CP}}$ becomes
small near $p=1$, which accounts for the larger finite-size and statistical
errors there.  Together with the short-ranged CMI, this provides no evidence
for either a mixed-state phase transition or a canonical-purification phase
transition at any finite $p<1$.

\subsection{The complete dephased point $p=1$}
\label{app:chern-endpoint}

\begin{figure*}[t]
 \centering
 \includegraphics[width=0.94\textwidth]
 {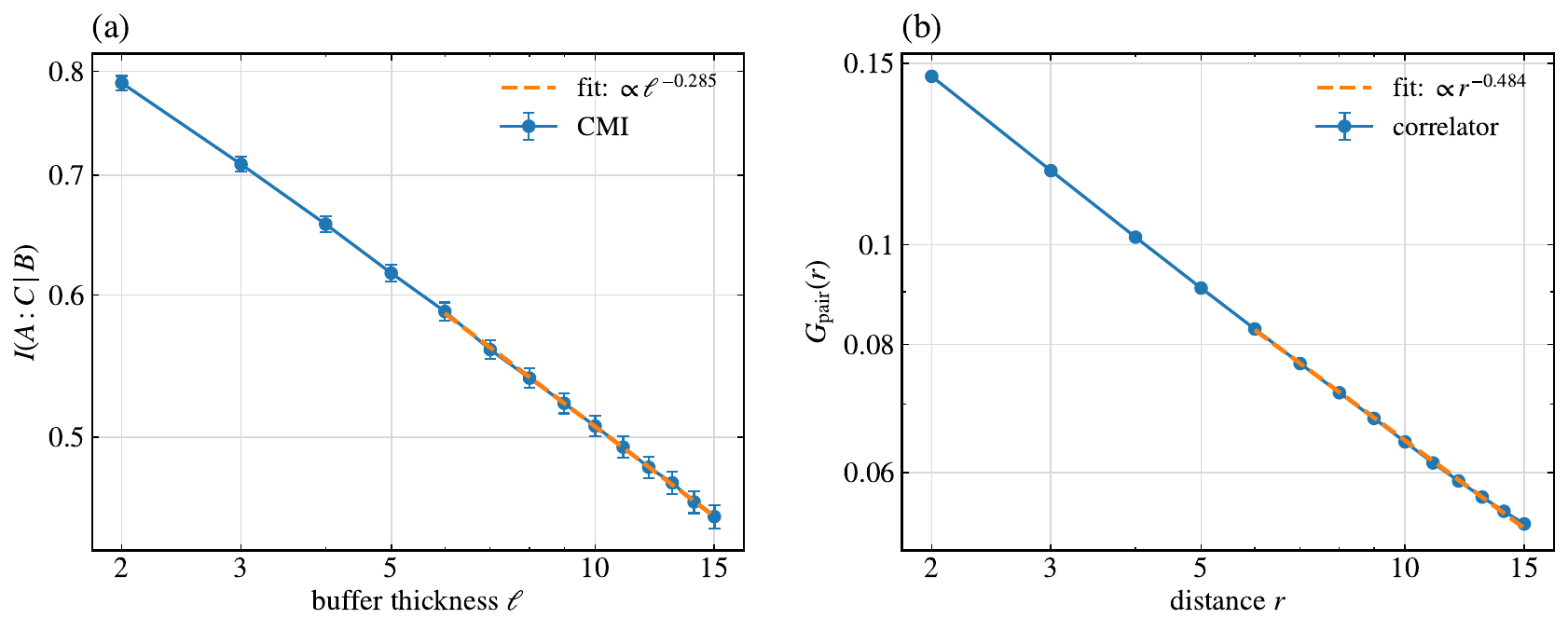}
 \caption{Fully measured endpoint.  The CMI (left) and the paired-copy
 correlator of the canonical purification (right) are shown with one-standard-
 error Monte Carlo uncertainties.  Solid curves are weighted power-law fits
 over separations $6$ through $15$; points from $2$ through $15$ are displayed.
 The algebraic behavior of both observables is consistent with a gapless
 endpoint.}
 \label{fig:app-chern-p1}
\end{figure*}

At $p=1$, every occupation number is recorded.  The flag is then locked to the post-measurement fermion configuration and can be suppressed. Therefore we have a fully classical state
\begin{equation}
    \rho_1 = \sum_{\mathbf{m}} P(\mathbf{m}) |\mathbf{m}\rangle\langle\mathbf{m}|.
\end{equation}
We compute the CMI again by sampling the measurement record, and find that the CMI to be power law decaying, see Fig.~\ref{fig:app-chern-p1}.

The connected density correlator of the canonical purification remains short-ranged. However, the state has a long-ranged pairing correlator defined in the following way. Within the locally paired
subspace spanned by $\ket{00}_i$ and $\ket{11}_i$, define
\begin{equation}
 \tau_i^+=\ket{11}_i\!\bra{00}_i,
 \qquad \tau_i^-=\ket{00}_i\!\bra{11}_i.
\end{equation}
Then
\begin{equation}
 G_{\mathrm{pair}}(r)=
 \bra{\sqrt{\rho_1}}\tau_{j+r}^{+}\tau_j^{-}
 \ket{\sqrt{\rho_1}}.
 \label{eq:app-pair-correlator}
\end{equation}
The pairing correlator can be computed through sampling the trajectories, too, through the estimator
\begin{equation}
G_{\mathrm{pair}}(r)
=
\mathbb E_{\boldsymbol m\sim P}\left[
\mathbf 1_{m_j=1,\,m_{j+r}=0}
\sqrt{
\frac{P(\boldsymbol m^{j\rightarrow j+r})}
     {P(\boldsymbol m)}
}
\right],
\label{eq:app-pair-estimator}
\end{equation}
where $\boldsymbol m^{j\rightarrow j+r}$ is obtained from
$\boldsymbol m$ by moving the fermion at $j$ to $j+r$. In practice, we
average the estimator over all choices of $j$ to reduce the statistical error.

We evaluated both quantities for the $L_x=L_y = 48$ system at flux $2\pi/4$ and filling $1/4$, using $10^4$ independent samples for
each size, see Figure~\ref{fig:app-chern-p1}. We fit that 
\begin{equation}
 I(A:C|B_{\ell})\simeq0.976\,\ell^{-0.285},
 \qquad
 G_{\mathrm{pair}}(r)\simeq0.197\,r^{-0.484}.
 \label{eq:app-endpoint-fits}
\end{equation}
Thus we conclude that the fully dephased Chern insulator has long range CMI and the canonical purification has long-range correlations, indicating a phase transition at $p=1$.

\end{document}